%% file: BRIE.tex
\documentclass[final,1p,times]{elsarticle}

\usepackage{amssymb}
\usepackage[export]{adjustbox}
\usepackage{caption}
\usepackage{booktabs}
\usepackage{subcaption}
\usepackage{url}
\usepackage{soul}
\usepackage{amsmath}
\usepackage{float}
\usepackage{xcolor}
\usepackage{microtype}

\journal{Information Fusion}

\begin{document}

\begin{frontmatter}



\cortext[corauthor]{Corresponding author.}


\title{Sustainable Transparency on Recommender Systems: Bayesian Ranking of Images for Explainability}


\author[inst1]{Jorge Paz-Ruza\corref{corauthor}}
\ead{j.ruza@udc.es}
\author[inst1]{Amparo Alonso-Betanzos}
\ead{amparo.alonso.betanzos@udc.es}
\author[inst1]{Bertha Guijarro-Berdiñas}
\ead{berta.guijarro@udc.es}
\author[inst1]{Brais Cancela}
\ead{brais.cancela@udc.es}
\author[inst1]{Carlos Eiras-Franco}
\ead{carlos.eiras.franco@udc.es}

\affiliation[inst1]{organization={Universidade da Coruña, CITIC},
            addressline={Campus de Elviña s/n}, 
            postcode={15008}, 
            state={A Coruña},
            country={Spain}}

\begin{abstract}

\input{0_Abstract}

\end{abstract}



\begin{keyword}
Machine Learning \sep Explainable Artificial Intelligence\sep Frugal AI \sep Dyadic Data\sep Explainable Recommendations\sep Recommender Systems

\PACS 07.05.Mh
\MSC[2020] 68T01
\end{keyword}

\end{frontmatter}


\input{secciones.tex}

\section{Acknowledgements}
\input{7_Acknowledgements}

\appendix
\input{A_Appendix}



\bibliographystyle{elsarticle-num} 
\bibliography{BRIE} 





\end{document}

%% file: 0_Abstract.tex
Recommender Systems have become crucial in the modern world, commonly guiding users towards relevant content or products, and having a large influence over the decisions of users and citizens. However, ensuring transparency and user trust in these systems remains a challenge; personalized explanations have emerged as a solution, offering justifications for recommendations. Among the existing approaches for generating personalized explanations, using existing visual content created by users is a promising option to maximize transparency and user trust. State-of-the-art models that follow this approach, despite leveraging highly optimized architectures, employ surrogate learning tasks that do not efficiently model the objective of ranking images as explanations for a given recommendation; this leads to a suboptimal training process with high computational costs that may not be reduced without affecting model performance. This work presents BRIE, a novel model where we leverage Bayesian Pairwise Ranking to enhance the training process, allowing us to consistently outperform state-of-the-art models in six real-world datasets while reducing its model size by up to 64 times and its CO${_2}$ emissions by up to 75\% in training and inference. 

%% file: secciones.tex
\input{1_Introduction}

\input{2_Background}

\input{3_Proposal}

\input{4_Experimental_Setup}

\input{5_Results}

\input{6_Conclusions}

%% file: 1_Introduction.tex
\section{Introduction}
\label{sec:introduction}


In recent years, the field of Artificial Intelligence (AI) has witnessed a significant rise in the development, research, and deployment of complex AI systems that can have a direct impact on users' and citizens' lives. From decision-making processes to personalized recommendations, they have become necessary components of numerous applications and fields, enhancing user experience and improving the communicative capabilities of institutions and companies. However, as these systems grow in complexity and influence on society, there is a pressing need to ensure their transparency and explainability, particularly in domains where user trust is paramount \cite{doshi2017accountability}.

The growing importance of explainability in AI systems during the past decade stems from the need to address the ``black box'' nature of many complex models \cite{ribeiro2016should}, as users often find it challenging to comprehend why an AI system produces a specific output, such as an item or product recommendation provided by a Recommender System. This lack of transparency has proved to lead to a diminished user experience, reduced trust from society in AI systems, and, in the worst case, potentially harmful outcomes \cite{hamm2023explanation}. This has popularly led to regulatory efforts to ensure transparency and accountability in AI systems. For instance, the European Union's General Data Protection Regulation (GDPR) mandates the right to explanation in data-treating systems, granting users insight into any automated decisions made by them \cite{gdpr}. Ethical-oriented guidelines, such as those published by the EU's High-Level Expert Group on Artificial Intelligence, emphasize transparency and accountability in AI \cite{eu-guidelines}. Additionally, the U.S. Federal Trade Commission (FTC) emphasized the need for explainable AI to prevent deceptive practices in a recent report \cite{ftc-report}. More recently, AI-specific legislation such as the EU'S AI Act has mandated AI systems to be designed in developed in such a way that their operator is sufficiently transparent to enable developers and users to interpret the system's output and use it appropriately \cite{hacker2023sustainable, Aiact}.

With respect to recommendations, the increasing reliance of businesses and users on AI-based Recommender Systems (RS) means that understanding and interpreting the recommendations provided to users has become crucial. Users expect personalized recommendations that align with their preferences, but they also desire explanations for why certain recommendations are made, e.g. due to a lack of quality of the recommendation or mistrust in the system; as such, explaining the outputs of RS not only enhances user trust, but also further enables users to make informed decisions, promoting a more interactive and engaging user experience \cite{zhang2020explainable}.

In the quest for explainability, one promising avenue is leveraging the visual content associated with AI inputs and outputs. Visual information, such as images, plays a vital role in human perception and decision-making processes; it has been explored before as a tool for explainability in other tasks \cite{zeiler2014visualizing}.  In the context of user-to-item recommendations, there exist many examples of visual-based explainability: Netflix uses a small selection of movie thumbnails crafted by experts to cater to different types of users \cite{netflix-research-2023}; Chen, Xu, et al. \cite{chen2019personalized} leverage user tastes from textual fashion reviews to highlight different image sections of the fashion item being recommended; Guesmi et al. \cite{guesmi2023justification} and Xu et al. \cite{xu2023does} generate word cloud images to visually explain what user or item characteristics prompted article or movie recommendations. Consequently, utilizing visual content as a means to explain recommendations can greatly enhance user understanding and satisfaction \cite{hendricks2016generating} or the effectiveness of industrial processes \cite{fikret2023transparent}.

Within the idea of visual explanations, our particular approach of interest is the use of \textit{user-uploaded images} to provide the explanations, avoiding a need for auxiliary information (e.g. text) to distil user tastes or a manual selection of creation of the images, thus yielding more organic, natural explanations and improving the self-sustainability and scalability of the explainable methodology. Compared to other visual-based approaches, which involve the generation of synthetic images, utilizing existing real images of the item recommended also provides a more implicit, self-contained and trustworthy explanation inside the recommendation pipeline, as the explanations will always be representative of the item's features.

\begin{figure}[]
    \centering
    \includegraphics[width=\textwidth]{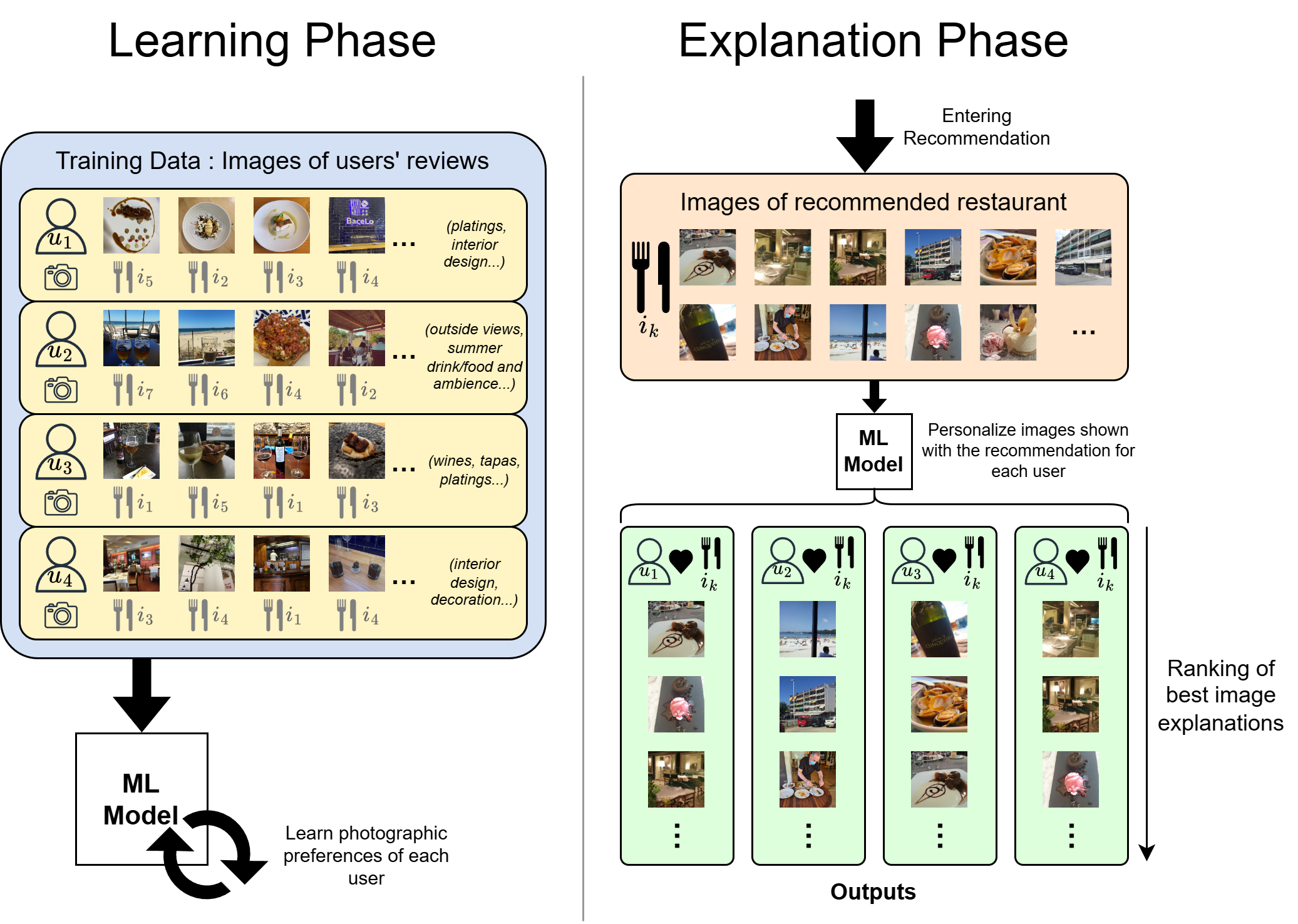}
    \caption{Overview of the use of user-uploaded images for visual-based explainability and personalization recommendation. Users upload images to justify their experiences with an item, so image features are good explanations for the user; a model can learn these features for each user (left half). Then, the appeareance of any entering recommendation can be personalized (here, with an image of the recommended restaurant) using the item's image that best reflects the learnt explanatory preferences of the user; the chosen image explanation will vary between users depending on their preferences (right side).}
    \label{fig:task}
\end{figure}

Leveraging user-uploaded images is a promising but particularly challenging task: in the case of textual contents, a textual explanation may be associated with multiple users or items (e.g. the phrase ``great visual effects'' in movie reviews), which can help in tackling data sparsity \cite{li2023relationship}. User-uploaded visual explanations (e.g. an image taken in a restaurant), however, are unique to each user and item, making it difficult for a model to distil user tastes. This challenge has been tackled by ELVis \cite{diez2020towards} and MF-ELVis \cite{paz2022sustainable}, which model and solve the explanation problem as a Learning to Rank (LTR) authorship task, predicting the best explanations of a \textit{(user, item)} recommendation as a ranking of the images of the item most likely to have been authored by the user. Figure \ref{fig:task} illustrates this approach, which is also the object of research of this work.

While explainability is crucial, it should not come at the expense of sustainability and model compactness. AI and Deep Learning models are known for their high computational cost, carbon emissions and model sizes; the concept of sustainable or ``green'' AI has become a popular research topic in the past years \cite{schwartz2020green}. These concerns have highlighted the need to balance model efficiency and explainability \cite{strubell2019energy} \cite{cheng2018model}. ELVis first provided explainability to an RS by leveraging user-uploaded images, but used a Deep Learning-oriented architecture with a cost comparable to the RS itself. Although MF-Elvis optimized the architecture of the model with a simpler yet adequate Matrix Factorization approach, both models still need a costly training process and a large model size to obtain good results. We believe one potential culprit of this computational cost is the inadequacy of the training objective used by both ELVis and MF-ELVis, which model the task of ranking the best image explanations as a surrogate binary classification of image authorship, which may be far from optimal. 

In this work, we introduce BRIE (Bayesian Ranking of Images for Explainability), a novel model designed to explain the outputs of Recommender Systems using existing images uploaded by the users themselves. BRIE addresses the challenges mentioned above not only surpassing the performances of the state-of-the-art models, ELVis and MF-ELVis, but also achieving drastically higher efficiency and improving model compactness. Unlike prior state-of-the-art approaches, BRIE models a more adequate training objective closer to the final Learning to Rank task, allowing it to explain recommendations better while reducing the cost of training and inference in terms of model size, training time and carbon emissions. 

The primary contributions of this research are threefold:

\begin{itemize}
    \item We introduce, design and implement BRIE, a novel model that leverages existing user-uploaded images to provide visually interpretable explanations for Recommender System outputs. BRIE improves existing approaches by adopting a more adequate learning objective closer to the final use of the model, while maintaining a highly optimized and simple architecture.
    \item We show that BRIE outperforms the state-of-the-art models of the task, achieving higher performance in six real-world datasets for image-driven explanation of recommendations. 
    \item We demonstrate that BRIE helps to address the inherent issue of sustainability in Explainable AI by minimizing carbon emissions, training time and particularly model size, ensuring that an improvement in explainability can be achieved without sacrificing the efficiency of the explained model. 
\end{itemize}

The structure of the paper is organized as follows. In Section \ref{sec:background}, we introduce the visual explanations ranking task and discuss existing state-of-the-art approaches and their limitations. Section \ref{sec:proposal} presents BRIE, our proposed algorithm that addresses these limitations through a better learning goal and use of training data. Section \ref{sec:experimentalsetup} describes the experimental setup, including real-world datasets, performance and efficiency evaluation metrics, and implementation details. In Section \ref{sec:results}, we present and discuss the performance of BRIE compared to state-of-the-art approaches and baselines. Finally, Section \ref{sec:conclusions} summarizes the contributions and implications of BRIE in terms of explainability, efficiency, and sustainability.

%% file: 2_Background.tex
\section{Background}
\label{sec:background}
\subsection{Problem Formulation}

In this subsection, we provide a formalization of the problem at hand: predicting the most suitable visual contents (photographs or images) to explain item recommendations to users made by any Recommender System or preceding information filtering algorithm.

Let $\mathcal{U}$ denote the set of users, $\mathcal{I}$ denote the set of items, and $\mathcal{P}$ denote the set of photographs of the items uploaded by users. The objective is to predict the most suitable existing photographs that effectively can explain the recommendation of an item to a user. The available information to do so (the ``historical interactions'' between $\mathcal{U}$, $\mathcal{I}$ and $\mathcal{P}$) forms a set of \textit{triads} as $\mathcal{D} = \{(u, i, p)\}$, where each triad represents a user $u \in \mathcal{U}$ interacting with an item $i \in \mathcal{I}$ by uploading the photograph $p \in \mathcal{P}$; reasonably, $p$ likely includes features of $i$ that are relevant to the user $u$, and therefore would be a good explanation of the item $i$ to user $u$. 

Utilizing user-uploaded photographs as explanations for items introduces certain considerations that must be addressed. Specifically, in most data contexts, each photograph is deemed a plausible explanation solely for the item it represents or is associated with, such as photographs of a hotel room or images found in product reviews. Also notably, $\mathcal{D}$ contains only one triad for each photo $p \in \mathcal{P}$, although a photo could be a relevant explanation for multiple users.

Therefore, leveraging user-uploaded item images as visual explanations poses two main challenges: 1) extracting meaningful latent information from photographs, given their limited association with multiple users and items, and 2) addressing data sparsity in real-world datasets where the available set of photographs $\mathcal{P}$ may be insufficient compared to the set of users $\mathcal{U}$, and no additional information (e.g. demographic variables) is available.

However, it is important to note that any photograph $p \notin \mathcal{P}_i$, i.e., any photograph that does not correspond to item $i$, can be readily disregarded as a plausible explanation for any recommendation of item $i$ to a particular user. This simplifies the formulation of the task by narrowing the focus to only the photographs associated with the specific item under consideration.

Consequently, we can focus solely on the interaction matrix between users and photographs, denoted as $\mathbf{R} \in \mathbb{R}^{|\mathcal{U}| \times |\mathcal{P}|}$, where $|\mathcal{U}|$ and $|\mathcal{P}|$ denote the cardinalities of the user and photograph sets, respectively. Each element $\mathbb{R}_{up}$ of $\mathbf{R}$ represents the relevance or preference score that user $u$ assigns to photograph $p$. Our goal is to determine the optimal ranking function $f: \mathcal{U} \times \mathcal{I} \times \mathcal{P}_i \rightarrow \mathbb{R}$, which can assign a score to each photograph $p \in \mathcal{P}_i$ for a given user-item pair $(u, i)$. This ranking function reflects the probability that user $u$ would have ``authored" photograph $p$ if they had interacted with item $i$. Thus, the higher the score assigned by $f$, the more suitable the photograph is as an explanation for the recommendation.

In particular, in order to identify the top explanation, we can find the photograph $p^*$ that maximizes the ranking function $f$ for a given user-item pair $(u, i)$:

\begin{equation}
p^* = \arg\max_{p \in \mathcal{P}_i} f(u, i, p)
\end{equation}

In other words, $p^*$ represents the photograph with the highest score according to $f$, indicating the most probable explanation for the recommendation of item $i$ to user $u$. 

As an explainability task, this problem assumes that the item $i$ being recommended is already given in a previous step by a Recommender System or any other information filtering system that matches users with items, and therefore focuses on finding a ranking function $f$ that can capture nuanced user preferences and photograph characteristics to provide adequate explanations of any given user-item pair $(u,i)$ while remaining disentangled from the overarching user-item pairing system.

\subsection{ELVis and MF-ELVis}

\begin{figure}[H]
    \begin{subfigure}[t]{0.45\textwidth}
    \begin{center}
    \textbf{ELVis} \\
    \vspace{.2cm}
    \includegraphics[width=\textwidth, valign=t]{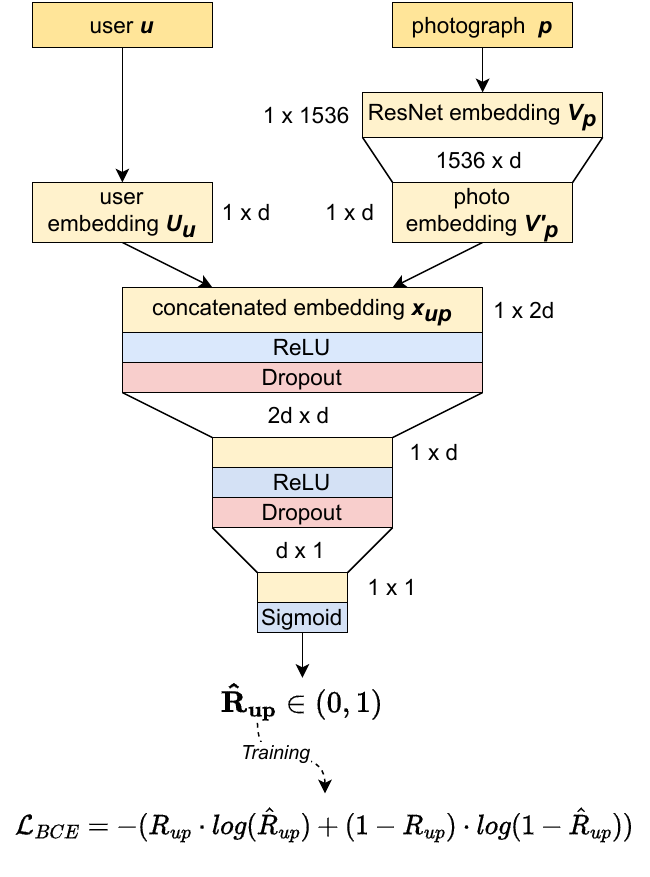}
    \end{center}
    \end{subfigure}
    \hspace{1cm}
    \begin{subfigure}[t]{0.45\textwidth}
    \begin{center}
    \textbf{MF-ELVis} \\
    \vspace{.2cm}
    \includegraphics[width=\textwidth, valign=t]{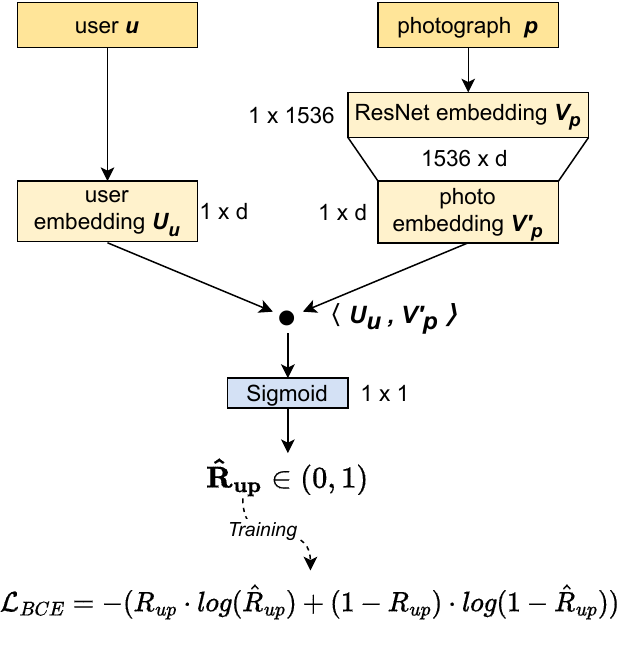}
    \end{center}
    
    \end{subfigure}
    
    \caption{Network topologies of ELVis and MF-ELVis, the two existing models for visual explainability through user-uploaded images in Recommender Systems. Both models take as inputs the IDs of user $u$ and photograph $p$, and output the predicted authorship \(\mathbf{\hat{R}}_{up}\), i.e. how good is $p$ as an explanation of a recommendation of its associated item for user $u$. During training, both models minimize BCE loss $\mathcal{L}_\text{BCE}$ based on the predicted and real authorship \(\mathbf{\hat{R}}_{up}\) and \(\mathbf{R}_{up}\).}
    \label{fig:ELVisandMFELVis}
\end{figure}

ELVis and MF-ELVis are the two existing models current state-of-the-art approaches for visual explainability through user-uploaded images, a novel approach for explainability in Recommender Systems. In this subsection, we describe the functioning of these models and provide insights into their design choices. Fig. \ref{fig:ELVisandMFELVis} depicts in detail the network topologies of both models.

\subsubsection{ELVis} Díez et al. \cite{diez2020towards} propose in ELVis to capture the underlying patterns and preferences in the interactions between users and photographs by representing the interaction matrix $\mathbf{R}$ as two latent matrices: the matrix of latent user preferences $\mathbf{U}$ and the matrix of latent photograph features $\mathbf{V'}$. 

ELVis generates a latent user vector $\mathbf{U}_u \in \mathbb{R}^d$ based on the user's ID, and obtains a latent photograph vector $\mathbf{V}_p' \in \mathbb{R}^d$ by projecting the ResNet-V2 \cite{szegedy2017inception} embedding $\mathbf{V}_p \in \mathbb{R}^{1536}$  of the original photograph into the latent factor subspace. The authors then propose to concatenate both vectors as $\mathbf{x}_{up}$, which is then passed through a multi-layer perceptron (MLP) $g$ to apply a series of non-linear transformations, such that

\begin{equation}
\mathbf{\hat{R}}_{up} =  \sigma(g([ \mathbf{U}_u; \mathbf{V}_p']))
\end{equation}

\noindent where $\sigma$ is the sigmoid activation function (see Fig. \ref{fig:ELVisandMFELVis}), and the output $\mathbf{\hat{R}}_{up} \in [0,1]$ is the predicted preference score for the (user, photograph) pair, or in other words, the adequacy of photograph $p$ as an explanation of its associated item for user $u$. 

The rationale behind ELVis's design is based on using a learned similarity function, following the popular Neural Collaborative Filtering architecture \cite{he2017neural}. The principle is that MLPs are general function approximators, so they should be -in theory- equally or more expressive than a fixed similarity function, such as a cosine similarity or dot product, and should capture more nuanced latent user preferences. However, the usage of an MLP comes at the expense of a very high computational cost overhead in both training and inference, motivating later approaches to abandon the usage of deep architectures.

\subsubsection{MF-ELVis}
A later publication suggests MF-ELVis \cite{paz2022sustainable} to simplify the model architecture while maintaining performance. Similar to ELVis, MF-ELVis learns the latent user preference matrix $\mathbf{U}$ and item feature matrix $\mathbf{V'}$. However, instead of an MLP, MF-ELVis uses a fixed similarity function, the inner product, to compute the similarity between the corresponding user vector $\mathbf{U}_u$ and photograph vector $\mathbf{V}_p'$. For a given (user, photograph) pair, the preference score is calculated as 

\begin{equation}
\mathbf{\hat{R}}_{up} =  \sigma\langle \mathbf{U}_u, \mathbf{V}_p'\rangle
\end{equation}

The rationale for adopting the inner product in MF-ELVis is based on findings presented by Rendle et al. \cite{rendle2020neural}. The study suggests that, in similarity-related tasks, the classic inner product can achieve performance comparable to an MLP while simplifying the model architecture. Leveraging the inner product operation, MF-ELVis enhanced the time and carbon emissions required to train the model by simplifying the model architecture. However, it utilized a much higher number of latent factors $d$ to maintain competitiveness with ELVis, increasing the overall model size; this suggested that under the training protocols ELVis and MF-ELVis use, the model architecture could not be further simplified without losing performance, but other parts of the training pipeline, such as the modelled learning objective and the use of training data, could be suboptimal and an aspect to be improved.

\subsection{Limitations of ELVis and MF-ELVis}
\label{sec:learninggoals}

One notable challenge in the context of explanation ranking tasks is that the set of historical interactions, denoted as $\mathcal{D}$, contains only positive samples, which are triads representing good explanations of items to users, such that $\mathbf{R}_{up} = 1$ . However, to train a model that can learn the ranking function $f$, it is necessary to incorporate samples that represent suboptimal explanations. The challenge then lies in determining an appropriate strategy for selecting negative samples.

Both ELVis and MF-ELVis address this issue by employing a uniform negative sampling and a positive sample replication, as shown in the left half of Figure \ref{fig:SAMPLING}, yielding a static training dataset 40 times larger than the original with a balanced number of positive and negative samples. Specifically, for each positive sample  $(u,i,p) \in \mathcal{D}$, 39 additional samples are generated and used during all the training: 19 extra replications of $(u,i,p)$,  10 negative triads $(u,i,p')$ (where $p'$ are from the same restaurant as $p$ but uploaded by a different user $u'$), and 10 negative triads $(u,i',p')$ (where $p'$ were uploaded by a different user $u'$ in a restaurant other than $i$). In the field of Positive-Unlabelled Learning, this random selection of negative samples is considered a naive yet straightforward method \cite{bekker2020learning}.

Another essential aspect of learning the ranking function is determining the learning objective and corresponding loss function used to guide the training process. ELVis and MF-ELVis model the explainability task as a binary ``authorship'' problem, formulating it as a binary classification and using binary cross-entropy (BCE) loss, which is commonly employed in classification tasks. The models are then evaluated using the original Learning to Rank ``explanation rankings'' framework. Although the choice of binary classification can lead to reasonable results, it may not be the optimal approach for Learning to Rank tasks. One limitation is that BCE loss does not consider the relative ordering or the degree of difference between instances within the positive and negative classes, but ranking tasks require capturing the notion of preference and accurately modelling the relative ordering of samples. In other words, it is essential to distinguish between good explanations that are preferable to others, rather than simply classifying them as positive or negative.

%% file: 3_Proposal.tex
\section{BRIE}
\label{sec:proposal}
In this section, we present BRIE (Bayesian Ranking of Images for Explainability), a novel model that builds upon MF-ELVis and addresses its limitations to surpass state-of-the-art performance in serving user-uploaded visual content to explain recommendations, while drastically improving training and inference efficiency. 

Acknowledging the shortcomings of ELVis and MF-ELVis we introduced in Section \ref{sec:learninggoals}, BRIE seeks to improve the training pipeline by modelling a learning objective and use of training data that is more adequate for an explanation ranking task than using pure binary classification as a surrogate task. This approach allows BRIE to efficiently capture more nuanced user preferences and image characteristics, providing better, faster visual explanations for recommendations.  

\subsection{Modelling Explanation Ranking with BPR}

State-of-the-art approaches, like ELVis and MF-ELVis, utilize binary classification as a surrogate task to model the Learning to Rank scenario required to select the most adequate visual explanation of an item for a given user. Instead, BRIE uses the Bayesian Pairwise Ranking (BPR) \cite{rendle2012bpr} algorithm to learn the ranking function $f$. BPR is tailored to optimize model weights by maximizing the probability of correctly ranking positive ``good'' instances higher than negative ``bad'' instances. 

The assumption for defining what is a positive example (``good instances'') and a negative example (``bad instances'') is similar to the one defined by ELVis and MF-ELVis:

\begin{itemize}
    \item Regarding the positive examples, because the images of items uploaded by a given user $u$ (i.e. $\mathcal{P}_u$)  are representative of what $u$ uses to justify and personalize its reviews, then all images $p \in \mathcal{P}_u$ can be considered positive examples for that user. For any user $u$, we also can reasonably assume smoothness in the characteristics of images inside $\mathcal{P}_u$: across different restaurants and reviews, the user's images will generally focus on the same feature(s) of the items (e.g. a user that likes to upload images the interior design in each restaurant to justify their reviews).
    \item Concerning the negative examples, we maintain the simple, naive assumption made by ELVis and MF-ELVis, to better focalize the contributions of this work: any image not originally uploaded by $u$ may be selectable as a negative example for that user. For the sake of simplicity, this deliberately ignores that other images not originally uploaded by the user may share the explanatory features of the user's own images, i.e. they could be possible positive examples for $u$. As we introduced in Section \ref{sec:learninggoals}, this situation is common to data types where no real negative examples are known, so the full design of a PU Learning \cite{bekker2020learning} technique to tackle this complex issue is left as future work.

\end{itemize}

Although we hold similar assumptions for the labelling training examples, the policy for sampling training examples in BRIE differs from the one ELVis and MF-ELVis utilize for their binary classification learning, described in Section \ref{sec:learninggoals}. At the start of each training epoch, BRIE constructs an ``extended'' training set $\mathcal{D^+} = {(u,i,p,p_{neg})}$ by uniformly sampling, for each $(u,i,p) \in \mathcal{D}$, a photograph $p_{neg}$ that is not part of $\mathcal{P}_u \subset \mathcal{P}$, i.e. a photograph from a purely random item and not taken by user $u$.

\begin{figure}[h]
    \centering
    \includegraphics[width=.8\textwidth]{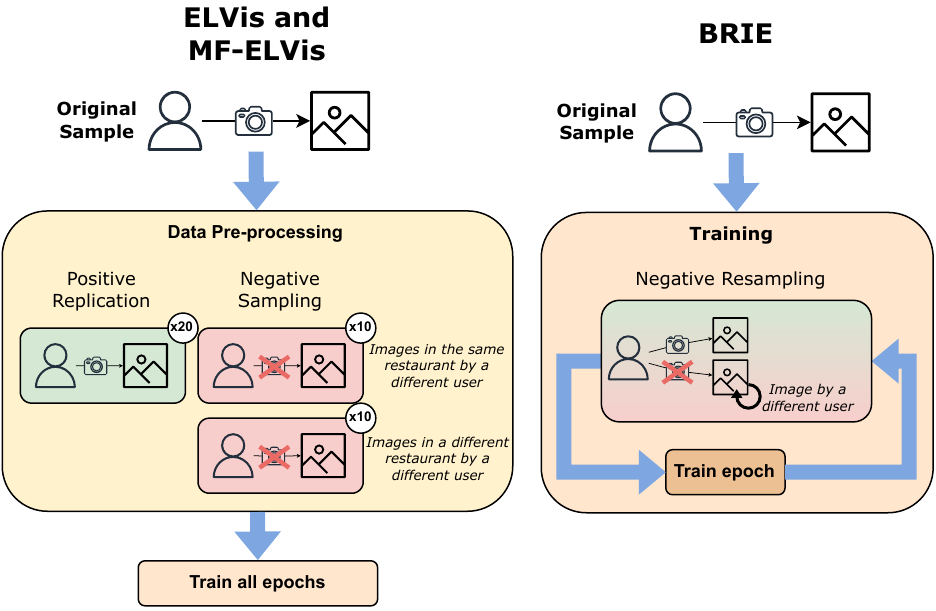}
    \caption{Negative sampling strategies of ELVis and MF-ELVis (left) versus BRIE (right), used for each original sample $(u, i, p)$ in the set of historical interactions $\mathcal{D}$.}
    \label{fig:SAMPLING}
\end{figure}

Our method is simpler than the pre-training selection used by ELVis and MF-ELVis, allows the regeneration of negative samples between epochs, and does not require storing all the pre-selected negative samples and replicated positive samples, unlike ELVis and MF-ELVis. Figure \ref{fig:SAMPLING} compares this selection policy with the one used in current state-of-the-art models. 

\begin{figure}[h]
    \centering
    \begin{subfigure}[t]{0.6\textwidth}
    \begin{center}
    \textbf{BRIE} \\
    \includegraphics[width=\textwidth, valign=t]{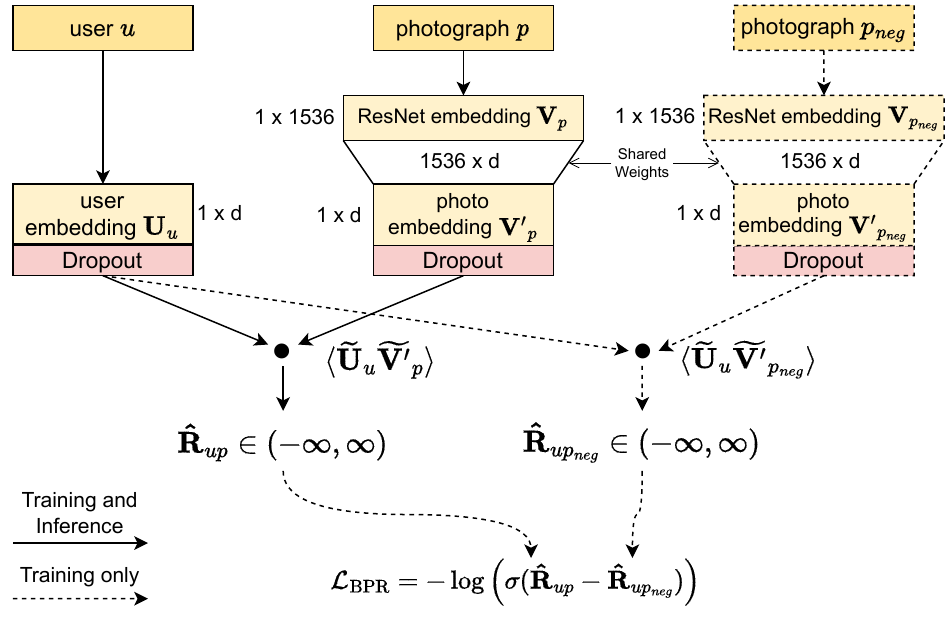}
    \end{center}
    \end{subfigure}
    \hfill
    
    \caption{Network topology of BRIE. While training, BRIE receives IDs of user $u$, a photograph $p$ taken by $u$, and a random photograph $p_{neg}$ assumed to not be adequate for $u$, and must maximize the difference of their predicted authorship probabilities (\(\mathbf{\hat{R}}_{up}-\mathbf{\hat{R}}_{up_{neg}}\)) to minimize BPR loss $\mathcal{L}_\textrm{BPR}$. During inference, BRIE receives IDs of user $u$ and photograph $p$, and outputs the predicted authorship \(\mathbf{\hat{R}}_{up}\), i.e. how good is $p$ as an explanation of a recommendation of its associated item for user $u$.}
    \label{fig:BRIE}
\end{figure}

If we formulate the BPR loss function over the extended dataset $\mathcal{D^+}$, then it can be defined as:

\begin{equation}
\mathcal{L}_{\textrm{BPR}} = -\sum_{(u, i, p, p_{neg}) \in \mathcal{D^+}} \log\left(\sigma(\mathbf{\hat{R}}_{up} - \mathbf{\hat{R}}_{up_{neg}})\right)
\end{equation}

\noindent where $\sigma$ represents the sigmoid function, $p$ is a photograph known to be authored by user $u$, and $p_{neg}$ is a photograph assumed to not be a good explanation for user $u$. Figure \ref{fig:BRIE} clarifies how the functioning of BRIE is designed to differ between training and inference: during inference, BRIE does not receive a second photograph $p_{neg}$, only a $(u, p)$ pair where $p$ is the photograph for which the predicted authorship $\mathbf{\hat{R}}_{up}$ needs to be calculated. 

By leveraging BPR as a basis for its training objective, BRIE learns a more adequate and effective ranking function that maximizes the probability of ranking better visual explanations over worse ones, rather than classifying them as ``good'' or ``bad'' explanations.

\subsection{Network Topology}
The foundation of BRIE's network topology takes advantage of the optimized architecture of MF-ELVis but, as seen in Figure \ref{fig:BRIE}, incorporates improvements and adaptations based on the changes introduced in the previous Section. 

Similarly to MF-ELVis, the latent user embedding $\mathbf{U}_u$ is extracted from the ID of user $u$, and the latent photograph embedding $\mathbf{V}_p'$ is obtained by projecting the ResNetv2 embedding of photograph $p$. During our initial experimentation we observed that, due to the improved adequateness of the BPR training objective and extended training set \(\mathcal{D}^+\), BRIE suffered overfitting if the dot product of $\mathbf{U}_u$ and $\mathbf{V}_p'$ was used directly to obtain $\mathbf{\hat{R}}_{up}$. To prevent this overfitting, dropout regularization is applied to $\mathbf{U}_u$ and $\mathbf{V}_p'$, resulting in modified embeddings $\widetilde{\mathbf{U}_u}$ and $\widetilde{\mathbf{V}_p'}$. The predicted authorship $\mathbf{\hat{R}}_{up} \in (-\infty,\infty)$ of the (user, photograph) pair can then be computed as:

\begin{equation}
    \mathbf{\hat{R}}_{up} = \langle\widetilde{\mathbf{U}_u} \widetilde{\mathbf{V}_p'}\rangle
\end{equation}

Note that, as BRIE doesn't employ a binary classification learning goal, there's no need to apply a sigmoid activation to map $\mathbf{\hat{R}}_{up}$ into the $[0,1]$ range.

%% file: 4_Experimental_Setup.tex
\section{Experimental Setup}
\label{sec:experimentalsetup}
\subsection{Datasets}
\label{sec:tripadvisor}

The datasets used in our experiments are derived from the original publication of ELVis by Diez et al. \cite{diez2020towards}, comprising photographs uploaded by users in the context of restaurant reviews on TripAdvisor, a popular hospitality establishment reviewing website. As far as our knowledge goes, this is the most diverse suite of datasets available for evaluating the use of user-uploaded images for visual-based RS explainability. We utilize public, real-world datasets from six different cities: Gijón, Barcelona, Madrid, New York, Paris, and London, in increasing number of photographs in the dataset, providing diverse contexts for evaluation in terms of socio-cultural characteristics, scale and user behaviours.

Each sample in the raw dataset consists of a triplet $(u, i, p) \in \mathcal{D}$, where $u$ represents the user ID, $i$ represents the restaurant ID, and $p$ represents the photograph ID. Table \ref{tab:datasets} summarizes key characteristics of each dataset, including the number of users, photographs, restaurants, and the overall sparsity of the datasets.

\begin{table}[htbp]
    \centering
    \caption{Basic statistics of each dataset.}
    \label{tab:datasets}
    \resizebox{.7\textwidth}{!}{%
    \begin{tabular}{lrrrr}
        \toprule
        \textbf{City} & \textbf{Users} & \textbf{Restaurants} & \textbf{Photographs} & \textbf{Photos/User}\\
        \midrule
        Gijón & 5,139 & 598 & 18,679 & 3.64\\
        Barcelona & 33,537 & 5,881 & 150,416 & 4.49\\
        Madrid & 43,628 & 6,810 & 203,905 & 4.67 \\
        New York & 61,019 & 7,588 & 231,141 & 3.79\\
        Paris & 61,391 & 11,982 & 251,636 & 4.10\\
        London & 134,816 & 13,888 & 479,798 & 3.56\\
        \bottomrule
    \end{tabular}
    }
\end{table}

For the sake of result reproducibility and comparability, we use the original train/validation/test partitions provided by the original authors. For each test case, we have one positive sample: a photograph $p$ that was not present in the training data but actually belongs to the user $u$. The remaining photographs $p' \in (\mathcal{P}_i \setminus \mathcal{P}_u $) of the same restaurant, contributed by different users, serve as negative samples. The goal of the model is to rank the positive sample (the user's own photograph) as high as possible among the negative samples, indicating its relevance as an explanation; this is a classic, popular approach to perform off-line evaluation in tasks involving implicit user-item interaction data \cite{shani2011evaluating}. Table \ref{tab:splits} shows an example partitioning of each of the datasets used in this work.

\begin{table}[h]
\caption{Example statistics (number of unique users and restaurants, and number of photographs) of a train-validation-test partitioning for each of the used datasets. Here, the number of photographs represents the number of positive examples in the partition, as well as the number of validation/test cases in the case of the validation/test sets.}
\label{tab:splits}
\resizebox{\textwidth}{!}{%
\begin{tabular}{lrrrrrrrrrrr}
\hline
                              & \multicolumn{3}{l}{\textbf{Train}}                                                                                       & \multicolumn{1}{l}{} & \multicolumn{3}{l}{\textbf{Validation}}                                                                                  & \multicolumn{1}{l}{} & \multicolumn{3}{l}{\textbf{Test}}                                                                                        \\ \cline{2-4} \cline{6-8} \cline{10-12} 
\multicolumn{1}{c}{\textbf{}} & \multicolumn{1}{c}{\textbf{Users}} & \multicolumn{1}{c}{\textbf{Restaurants}} & \multicolumn{1}{c}{\textbf{Photographs}} & \multicolumn{1}{l}{} & \multicolumn{1}{c}{\textbf{Users}} & \multicolumn{1}{c}{\textbf{Restaurants}} & \multicolumn{1}{c}{\textbf{Photographs}} & \multicolumn{1}{l}{} & \multicolumn{1}{c}{\textbf{Users}} & \multicolumn{1}{c}{\textbf{Restaurants}} & \multicolumn{1}{c}{\textbf{Photographs}} \\ \hline
Gijón                         & 5,139                              & 598                                      & 15,190                                   &                      & 428                                & 205                                      & 1,112                                    &                      & 1,023                              & 346                                      & 2,337                                    \\
Barcelona                     & 33,537                             & 5,881                                    & 120,221                                  &                      & 4,363                              & 2,033                                    & 10,453                                   &                      & 8,697                              & 3,211                                    & 19,742                                   \\
Madrid                        & 43,268                             & 6,810                                    & 162,487                                  &                      & 5,852                              & 2,301                                    & 14,276                                   &                      & 11,874                             & 3,643                                    & 27,142                                   \\
New York                      & 61,019                             & 7,588                                    & 179,093                                  &                      & 7,687                              & 2,571                                    & 17,222                                   &                      & 16,842                             & 4,135                                    & 34,826                                   \\
Paris                         & 61,391                             & 11,982                                   & 203,122                                  &                      & 7,417                              & 3,969                                    & 16,466                                   &                      & 15,242                             & 6,345                                    & 32,048                                   \\
London                        & 134,816                            & 13,888                                   & 384,528                                  &                      & 14,134                             & 5,121                                    & 31,828                                   &                      & 30,393                             & 8,097                                    & 63,442                                   \\ \hline
\end{tabular}%
}
\end{table}

It is worth noticing that, as datasets do not have any explicit information about each user (e.g. demographic), any model must learn the user's latent features before being able to select personalized explanations to that user (the classic ``cold start'' situation of user-item interaction data). To ensure a correct evaluation on this setting, Díez et al. \cite{diez2020towards} utilize a partitioning method common in tasks that deal with user-item interaction data \cite{10.1145/3383313.3418479}, which ensures the positive example hidden in each test case isn't a from new (cold-start) user, i.e. that all users are present in the train set; we refer readers to ELVIs' original publication \cite{diez2020towards} for more detailed information on the dataset extraction, characteristics, and partitioning.

\subsection{Compared Models}
We compare BRIE against the existing models for the task, as well as other baseline approaches, to evaluate its performance and effectiveness in visual-based explanation through user-uploaded images in Recommender Systems:

\begin{itemize}
\item \textbf{RND (Random)}: Assigns a random preference $\hat{\mathbf{R}}_{up}$ to each (user, photograph) pair. 

\item \textbf{CNT (Centroid)}: Estimates the preference of a (user, photograph) pair as the negative distance between the photograph ResNetv2 embedding $\mathbf{V}_p$ and the centroid of the ResNetv2 embeddings of the photographs belonging to user $u$. Mathematically, the estimated preference can be represented as $\hat{\mathbf{R}}_{up} = -|\mathbf{V}_p - \frac{1}{\mathcal{P}_u} \sum\limits_{p' \in \mathcal{P}_u} V_{p'}|$. 

\item \textbf{ELVis} \cite{diez2020towards}: Predicts the preference of a (user, photograph) pair applying a multi-layer perceptron $g$ to the concatenation of the user's latent embedding $\mathbf{U}_u$ and the photograph's latent embedding $\mathbf{V}_p'$ as input. The estimated preference can be represented as $\hat{\mathbf{R}}_{up} = \sigma(g( [\mathbf{U}_u; \mathbf{V}_p'])$.

\item \textbf{MF-ELVis} \cite{paz2022sustainable}: MF-ELVis predicts the preference of a (user, photograph) pair using a dot product between the user's latent embedding $\mathbf{U}_u$ and the photograph's latent embedding $\mathbf{V}_p'$. The estimated preference can be represented as $\hat{\mathbf{R}}_{up} = \sigma\langle \mathbf{U}_u, \mathbf{V}_p' \rangle$.
\end{itemize}

\subsection{Evaluation Metrics}

We evaluated the performance of the models using popular ranking metrics to assess their effectiveness in explanation ranking tasks, and compared their efficiency and compactness.

\subsubsection{Performance Evaluation Metrics}

Let $\mathcal{T}$ be the set of test cases constructed as described in Section \ref{sec:tripadvisor}, where the predictions for each test case form a ranking $f(u, i)$ that orders the best photographs from restaurant $i$ to explain a recommendation of the restaurant to user $u$. The performance evaluation metrics used in this work are as follows:

\begin{itemize}

  \item \textbf{Mean Recall at $k$ (MRecall@$\mathbf{k}$)}:
  \begin{equation}
        \textrm{MRecall@k} = \frac{1}{|\mathcal{T}|} \sum_{f(u, i) \in \mathcal{T}} \textrm{TP}_{f(u, i)}(k)
    \end{equation}

  where $\textrm{TP}_{f(u, i)}(k)$ is the number of true positive explanations at rank $k$ for the ranking $f(u, i)$.

  \item \textbf{Mean Normalized Discounted Cumulative Gain at $k$ (MNDCG@$\mathbf{k}$)}:
  \begin{equation}
  \textrm{MNDCG@k} = \frac{1}{|\mathcal{T}|} \sum_{f(u, i) \in \mathcal{T}} \frac{\textrm{DCG}_{f(u, i)}(k)}{\textrm{IDCG}_{f(u, i)}(k)}
  \end{equation}
  where $\textrm{DCG}_{f(u, i)}(k)$ is the discounted cumulative gain at rank $k$ for the ranking $f(u, i)$, and $\textrm{IDCG}_{f(u, i)}(k)$ is the ideal discounted cumulative gain at rank $k$ for the ranking $f(u, i)$.

  \item \textbf{Mean Area Under the ROC Curve (MAUC)}:
  \begin{equation}
  \textrm{MAUC} = \frac{1}{|\mathcal{T}|} \sum_{f(u, i) \in T} \textrm{AUC}_{f(u, i)}
  \end{equation}
  where $\textrm{AUC}_{f(u, i)}$ is the area under the ROC curve for the ranking $f(u, i)$.

  \item \textbf{Median Percentile (MedPerc)}:
  \begin{equation}
  \textrm{MedPerc} = \textrm{median}\left(\left\{\frac{R_{f(u, i)}}{N_{f(u, i)}} \times 100 \mid f(u, i) \in \mathcal{T}\right\}\right)
  \end{equation}
  where $R_{f(u, i)}$ is the position of the real positive sample (user's real photograph) in the ranking $f(u, i)$, and $N_{f(u, i)}$ is the total number of predictions for the ranking $f(u, i)$; therefore, the lower the percentile, the better the performance of the model.
\end{itemize}

As done by previous works, the computation of MRecall@$k$ and MNDCG@$k$ is performed with $k = 10$, and considering only the test cases that fulfil two conditions: First, the test case must contain at least 10 images to rank (1 positive example and at least 9 negative examples); this removes trivial test cases since, with $k=10$, any case with $\leq 10$ images would trivially return Recall@10=1 and NDCG@10$>$0. Second, the author user of the positive example in the test case must have least 10 positive examples (uploaded photographs) in the training set; removing these test cases that correspond to users in near-cold-start scenarios is common in user-interaction data to give a more reliable measure of ranking performance  \cite{shani2011evaluating, 10.1145/3383313.3418479}. On the other hand, when computing MAUC all test cases and users and considered, to have a measure the overall discriminative power of the models.

\subsubsection{Efficiency and Compactness Metrics}

In addition to the performance evaluation metrics, we also consider the following efficiency and compactness metrics:

\begin{itemize}
  \item \textbf{Evolution of Test MAUC with Training Time and Carbon Emissions}: This analysis helps understand the trade-off between training resource consumption and model effectiveness.

\item \textbf{Inference Time and Energy Cost}: This metric estimates the time and carbon emissions required to utilize the model in a practical use case, simulating a final production environment. To do this, we track the aforementioned criteria as we obtain the models' predictions for all the test cases in the test set of each dataset.

  \item \textbf{Impact of Number of Parameters in Test MAUC}: This comparison examines how the number of latent factors $d$ affects the performance of the models.
\end{itemize}

All training times and carbon emissions are tracked using Codecarbon \cite{schmidt2021codecarbon}.  

\subsection{Implementation Details}

\subsubsection{Framework and Machine Configuration}

For the sake of reproducibility and a fair evaluation, we implemented BRIE and re-implemented ELVis and MF-ELVis in a common Python framework using PyTorch-Lightning\footnote{\url{https://www.pytorchlightning.ai/}}; all models share the same infrastructure for data loading, computation of metrics, and any other processes not unique to the particular design of a specific model. We make this framework available in a public repository\footnote{\url{https://github.com/Kominaru/BRIE}}.

All experiments were conducted on a dedicated machine with 16GB RMA, an Intel Core i7-10700 processor, an NVIDIA RTX 2060 Super GPU and Windows 10 OS. Due to memory constraints produced by the large model sizes of ELVis and MF-ELVis, all models for Gijón, Barcelona, and Madrid were trained using 4 multiprocessing workers, models for New York and Paris utilized 2 workers, and models for London were trained without multiprocessing.

\subsubsection{Model Hyperparameters}

Originally, the authors of ELVis\footnote{\url{https://github.com/pablo-pnunez/ELVis}} \cite{diez2020towards} selected model hyperparameters through a grid search in the validation sets of each dataset, identifying a specific configuration per dataset; this approach was also shared by MF-ELVis\footnote{\url{https://github.com/Kominaru/tfg-komi}} \cite{paz2022sustainable}. In this work, we train and evaluate both models using the single best hyperparameters per dataset reported in their original publications.

Model hyperparameters for BRIE were selected through one random search in the validation set of Barcelona; this is the smallest available dataset that does not exhibit irregular or volatile performance evaluation, as found by Díez et al. \cite{diez2020towards}. We trained 25 models with random hyperparameters inside the following search space: latent factors ($d$) ranging from 4 to 1024, learning rate ($lr$) ranging from $10^{-3}$ to $10^{-5}$, and dropout ranging from 0 to 0.8. A fixed batch size of $2^{14}$ was used, and an early stopping policy was implemented with a patience $p$=5, a delta $\delta$=$10^{-3}$, and a maximum of 100 epochs while tracking validation MAUC during training, as suggested by the BPR publication \cite{rendle2012bpr}. All model parameters were initialized using a Xavier distribution \cite{glorot2010understanding} and optimized using Adam \cite{kingma2014adam}.

After a hyperparameter search on Barcelona's validation set, we trained on all datasets with the selected configuration, without early stopping. We selected the final hyperparameter combination based on both effectiveness and efficiency: considering only the configurations that consistently outperformed the State of the Art, we selected the one with the lowest inference emissions ($d$=64, $lr$=$10^{-3}$, dropout=0.75, 15 epochs), i.e. the one with the lowest $d$ (intuitively, the inference emissions of BRIE monotonically increase/decrease with larger/smaller $d$ values).

%% file: 5_Results.tex
\section{Results and Discussion}
\label{sec:results}

In this section, we compare BRIE with the existing state-of-the-art methods of the task and other baseline approaches using the experimental setup described in Section \ref{sec:experimentalsetup}. We evaluate the models' performance using standard metrics, assess their efficiency and sustainability in terms of time and energy consumption during training and inference, and investigate the impact of the number of latent factors on model performance.

\subsection{Performance comparison}

Table \ref{tab:performancesummary} presents the performance of all models on the six datasets using three evaluation metrics (MRecall@10, MNDCG@10, MAUC). BRIE consistently outperforms ELVis and MF-ELVis, the other two available models for the task, in all metrics and datasets, indicating its superior ability to provide explanations with user-uploaded images. On average, BRIE achieves a relative improvement of approximately 7.5\% over the second-best method in terms of MAUC, the most comprehensive metric. This suggests that BRIE's utilization of Bayesian Pairwise Ranking as a learning goal effectively captures nuanced preferences and interactions between users and photographs, and is as such a learning goal better suited for explanation ranking tasks than surrogate classification goals.

\begin{table}[H]
\caption{ Performance summary of all models across the six selected datasets. For each city and metric, the best model is boldfaced, and the second best model is underlined.}
\resizebox{\textwidth}{!}{%
\begin{tabular}{lrrrlrrrlrrr}
\hline
 &
  \multicolumn{3}{l}{\textbf{Gijón}} &
   &
  \multicolumn{3}{l}{\textbf{Barcelona}} &
   &
  \multicolumn{3}{l}{\textbf{Madrid}} \\ \cline{2-4} \cline{6-8} \cline{10-12} 
 &
  \multicolumn{1}{l}{MRecall@10} &
  \multicolumn{1}{l}{MNDCG@10} &
  \multicolumn{1}{l}{MAUC} &
   &
  \multicolumn{1}{l}{MRecall@10} &
  \multicolumn{1}{l}{MNDCG@10} &
  \multicolumn{1}{l}{MAUC} &
   &
  \multicolumn{1}{l}{MRecall@10} &
  \multicolumn{1}{l}{MNDCG@10} &
  \multicolumn{1}{l}{MAUC} \\ \hline
RND &
  0.373 &
  0.185 &
  0.487 &
   &
  0.409 &
  0.186 &
  0.502 &
   &
  0.374 &
  0.171 &
  0.499 \\
CNT &
  0.464 &
  0.218 &
  0.546 &
   &
  0.443 &
  0.219 &
  0.554 &
   &
  0.420 &
  0.203 &
  0.557 \\
ELVis &
  0.521 &
  0.262 &
  \underline{ 0.596} &
   &
  \underline{ 0.597} &
  \underline{ 0.327} &
  \underline{ 0.631} &
   &
  \underline{ 0.572} &
  \underline{ 0.314} &
  \underline{ 0.638} \\
MF-ELVis &
  \underline{ 0.538} &
  \underline{ 0.285} &
  0.592 &
   &
  0.557 &
  0.293 &
  0.596 &
   &
  0.528 &
  0.279 &
  0.601 \\
BRIE &
  \textbf{0.607} &
  \textbf{0.333} &
  \textbf{0.643} &
   &
  \textbf{0.630} &
  \textbf{0.368} &
  \textbf{0.663} &
   &
  \textbf{0.612} &
  \textbf{0.348} &
  \textbf{0.673} \\ \hline
\multicolumn{3}{l}{} &
  \multicolumn{1}{l}{} &
   &
  \multicolumn{2}{l}{} &
  \multicolumn{1}{l}{} &
   &
  \multicolumn{2}{l}{} &
  \multicolumn{1}{l}{} \\ \hline
 &
  \multicolumn{3}{l}{\textbf{Newyork}} &
   &
  \multicolumn{3}{l}{\textbf{Paris}} &
   &
  \multicolumn{3}{l}{\textbf{London}} \\ \cline{2-4} \cline{6-8} \cline{10-12} 
 &
  \multicolumn{1}{l}{MRecall@10} &
  \multicolumn{1}{l}{MNDCG@10} &
  \multicolumn{1}{l}{MAUC} &
   &
  \multicolumn{1}{l}{MRecall@10} &
  \multicolumn{1}{l}{MNDCG@10} &
  \multicolumn{1}{l}{MAUC} &
   &
  \multicolumn{1}{l}{MRecall@10} &
  \multicolumn{1}{l}{MNDCG@10} &
  \multicolumn{1}{l}{MAUC} \\ \hline
RND &
  0.374 &
  0.168 &
  0.502 &
   &
  0.459 &
  0.209 &
  0.502 &
   &
  0.342 &
  0.155 &
  0.500 \\
CNT &
  0.431 &
  0.217 &
  0.563 &
   &
  0.499 &
  0.245 &
  0.557 &
   &
  0.400 &
  0.200 &
  0.562 \\
ELVis &
  \underline{ 0.553} &
  \underline{ 0.304} &
  \underline{ 0.637} &
   &
  \underline{ 0.643} &
  \underline{ 0.352} &
  \underline{ 0.630} &
   &
    0.530 &
  \underline{ 0.293} &
  \underline{ 0.629} \\
MF-ELVis &
  0.516 &
  0.276 &
  0.602 &
   &
  0.606 &
  0.323 &
  0.596 &
   &
  \underline{ 0.531} &
  0.267 &
  0.597 \\
BRIE &
  \textbf{0.598} &
  \textbf{0.341} &
  \textbf{0.677} &
   &
  \textbf{0.669} &
  \textbf{0.391} &
  \textbf{0.666} &
   &
  \textbf{0.563} &
  \textbf{0.318} &
  \textbf{0.665} \\ \hline
\end{tabular}%
}
\label{tab:performancesummary}
\end{table}

While ELVis aims to achieve high performance by leveraging the expressiveness of the MLP architecture, it falls short in comparison to BRIE. In contrast, MF-ELVis simplifies the model by using the inner product as a fixed similarity function between user and photograph vectors; while this approach accelerates training and reduces carbon emissions, it compromises the model's expressiveness compared to ELVis and increases overfitting. As a result, MF-ELVis lags behind ELVis in performance and both methods fall behind BRIE, as they use binary classification as a surrogate learning goal, which is suboptimal for modelling the explanation ranking task. 

\begin{figure}[H]
  \centering
  \textbf{Median Percentile}

  \begin{subfigure}{0.48\textwidth}
    \includegraphics[width=\linewidth]{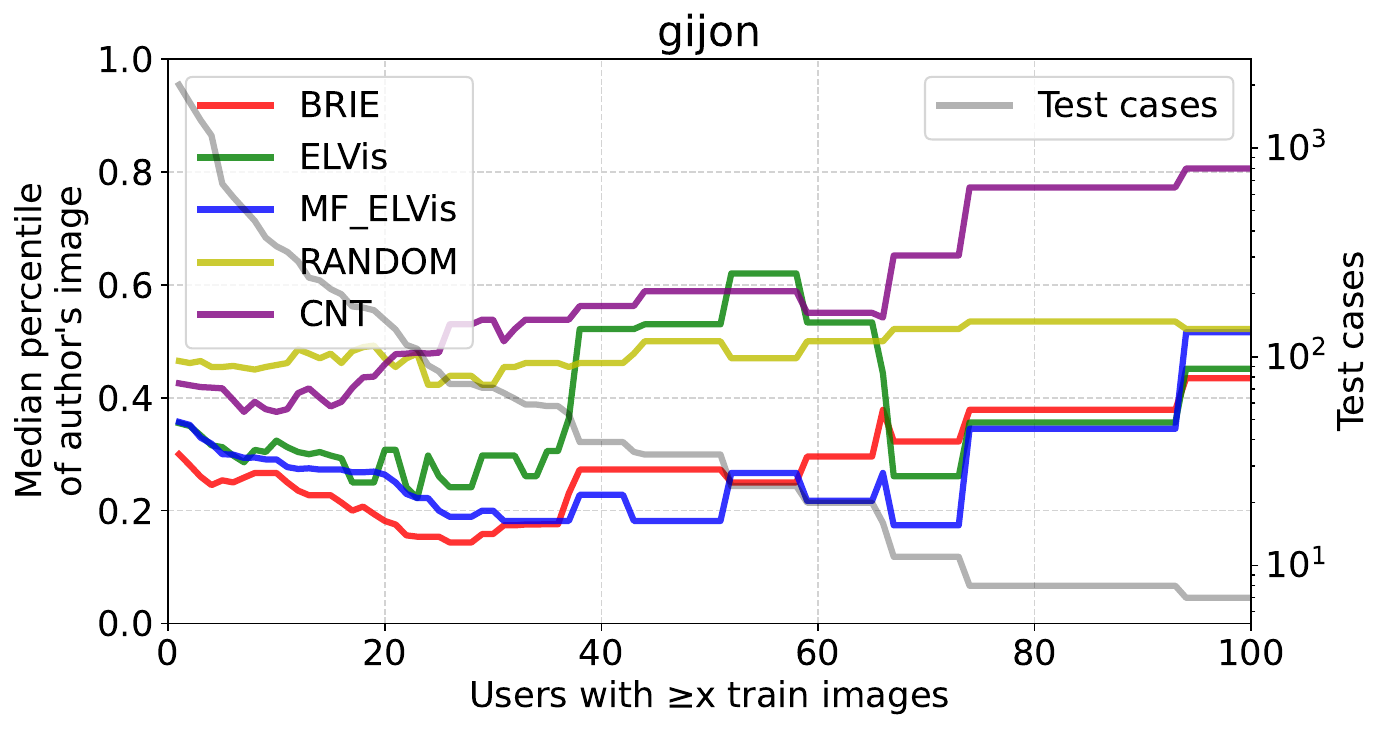}
  \end{subfigure}
  \hfill
  \begin{subfigure}{0.48\textwidth}
    \includegraphics[width=\linewidth]{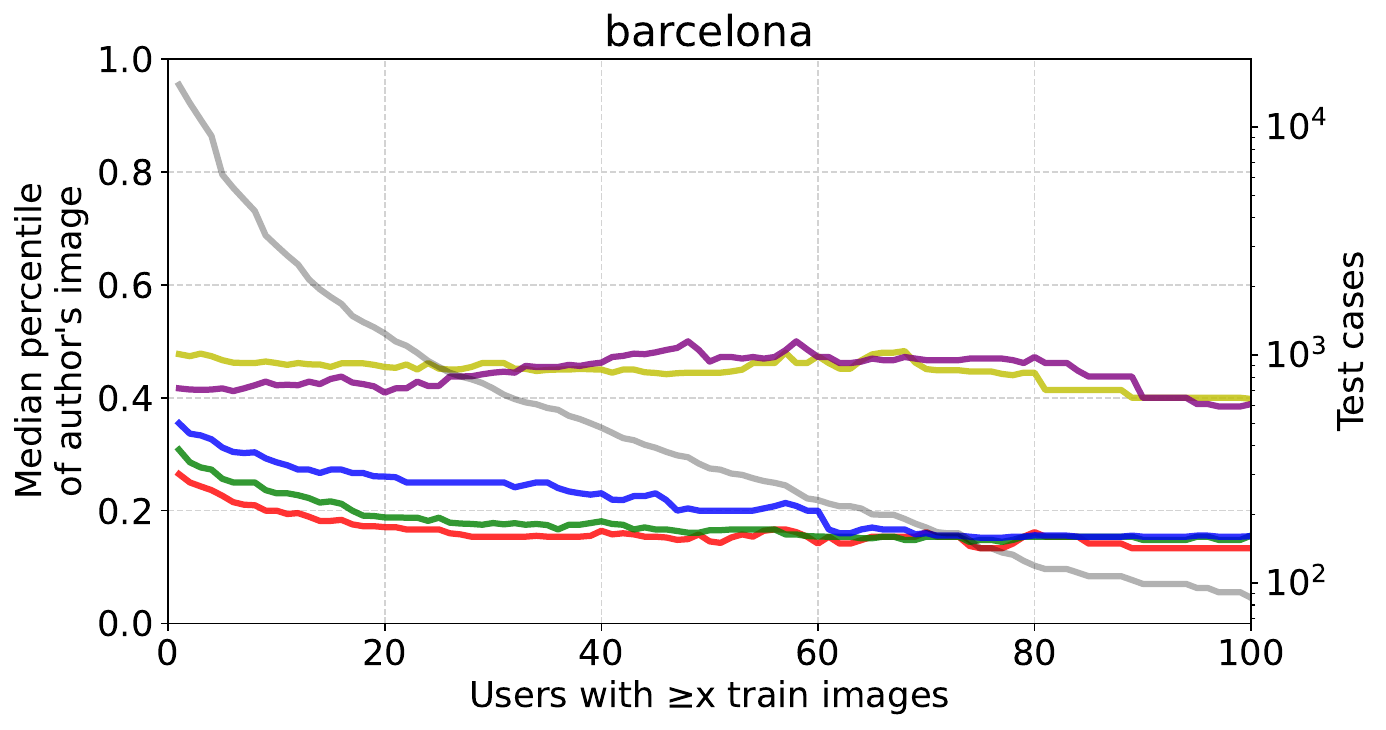}
  \end{subfigure}

  \begin{subfigure}{0.48\textwidth}
    \includegraphics[width=\linewidth]{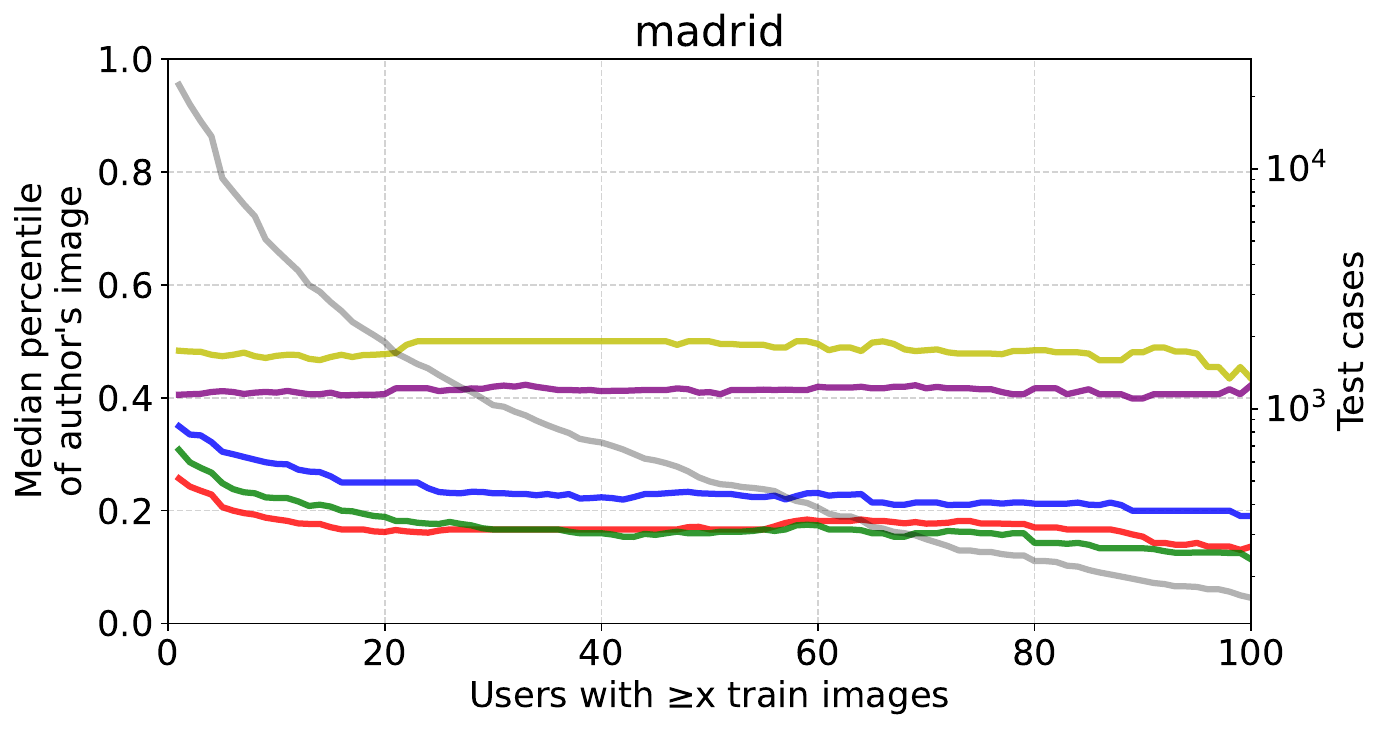}
  \end{subfigure}
  \hfill
  \begin{subfigure}{0.48\textwidth}
    \includegraphics[width=\linewidth]{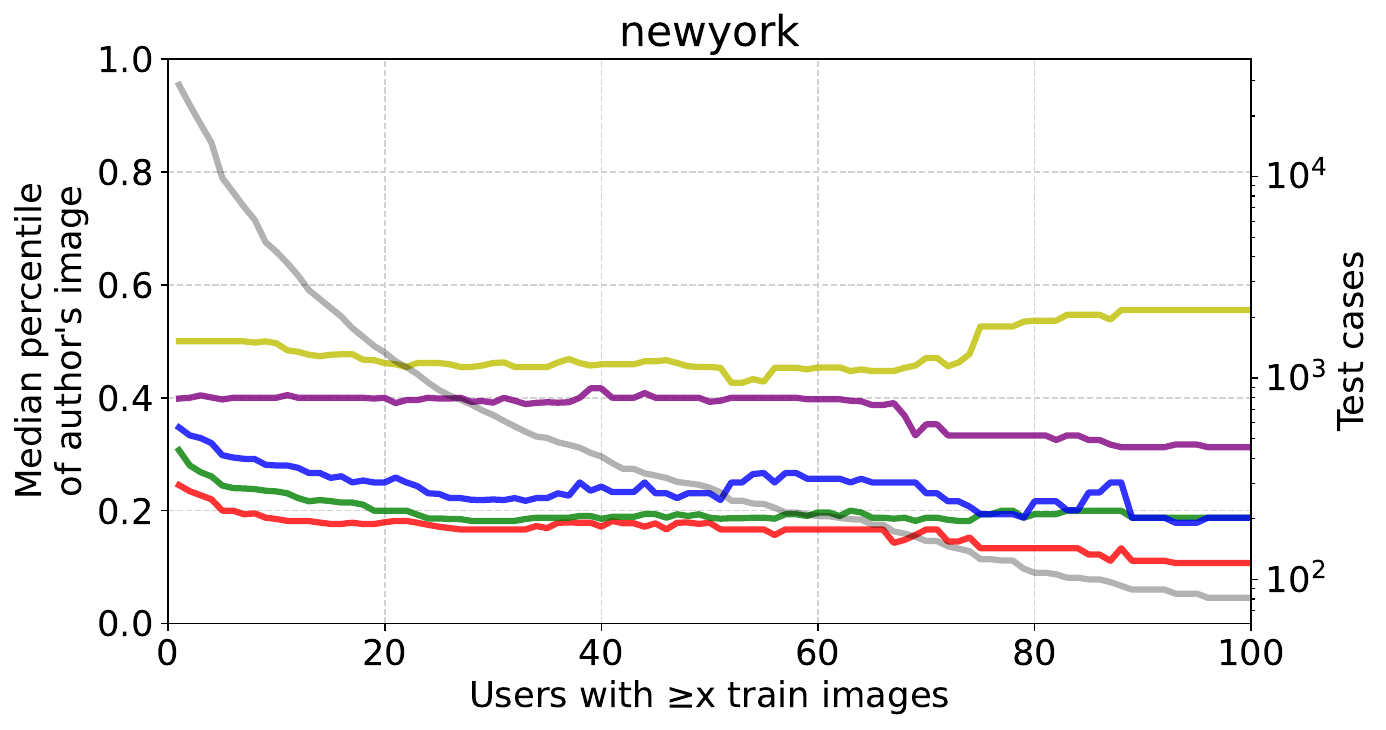}
  \end{subfigure}

  \begin{subfigure}{0.48\textwidth}
    \includegraphics[width=\linewidth]{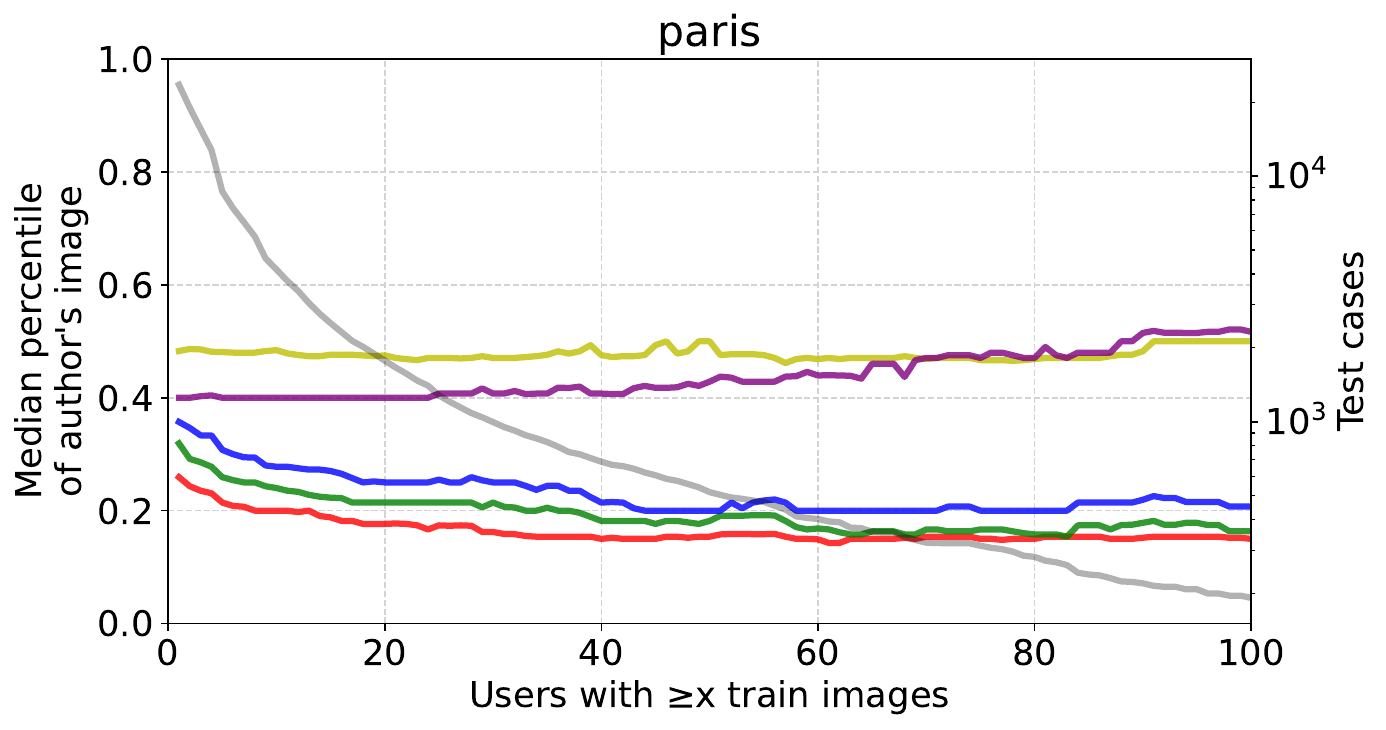}
  \end{subfigure}
  \hfill
  \begin{subfigure}{0.48\textwidth}
    \includegraphics[width=\linewidth]{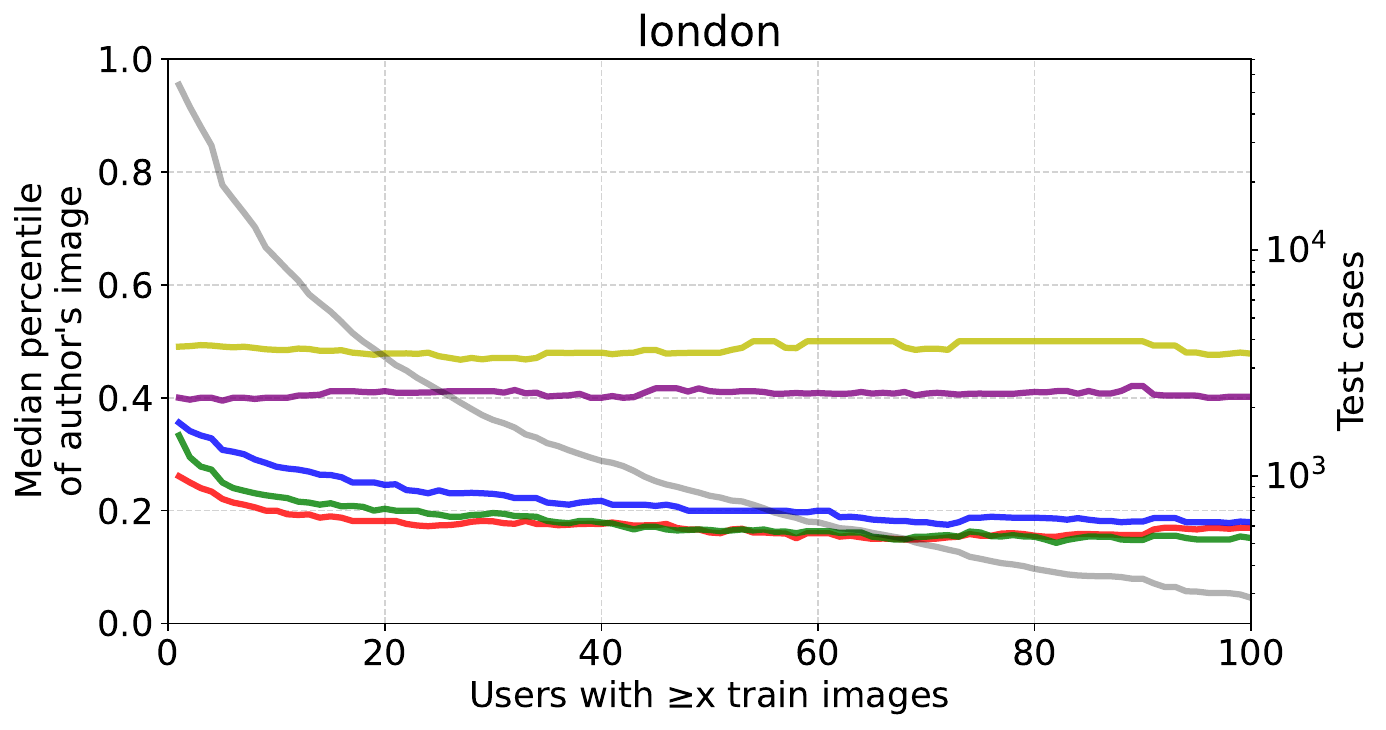}
  \end{subfigure}

  \caption{Effect of the amount of information per user on the median percentile (lower is better) of all models. Each subfigure displays the median percentile (y-axis) as a function of the minimum activity threshold (x-axis), i.e. the minimum required number of photographs by the user present in the training set. The count of available test cases for each minimum activity threshold is also provided as an aid in assessing the statistical significance of the results.}
  \label{fig:median_percentile}
\end{figure}

The median percentile analysis, as shown in Figure \ref{fig:median_percentile}, provides insights into the models' performance with varying user information availability. Generally, higher activity thresholds (i.e. more available information about each user) lead to improved performance for all models (except Random and CNT). While it is challenging to compare BRIE and ELVis in scenarios with abundant user information, BRIE consistently outperforms all models when user information is scarce. This suggests BRIE's strength in providing explanations with limited user data, making it a robust choice for diverse scenarios.

In summary, BRIE consistently outperforms ELVis, MF-ELVis, and basic baselines in all analyzed metrics and datasets. The ability to capture nuanced preferences and interactions, coupled with an efficient and well-suited learning goal, contributes to BRIE's superior performance even in scenarios where user information is scarce.

\subsection{Effiency and sustainability comparison}

In this subsection we compare the efficiency and sustainability of BRIE, ELVis, and MF-ELVis by comparing the evolution of AUC versus training time and CO$_2$ emissions for all models and datasets, as well as inference times and CO$_2$ emissions. Both training and inference are performed using GPU acceleration, and the final models used for reporting the results in Table \ref{tab:performancesummary} are considered. 

\begin{figure}[htbp]
    \begin{center}
        \textbf{Training Time vs. Test AUC}
    \end{center}
  \centering

  \begin{subfigure}[b]{0.32\textwidth}
    \centering
    \includegraphics[width=\textwidth]{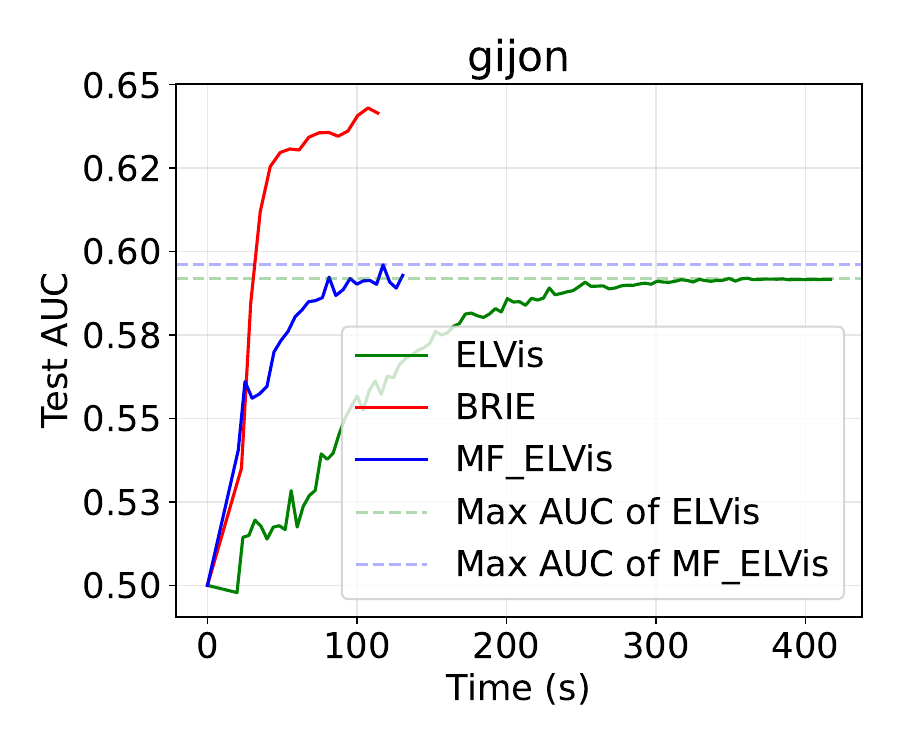}
  \end{subfigure}
  \begin{subfigure}[b]{0.32\textwidth}
    \centering
    \includegraphics[width=\textwidth]{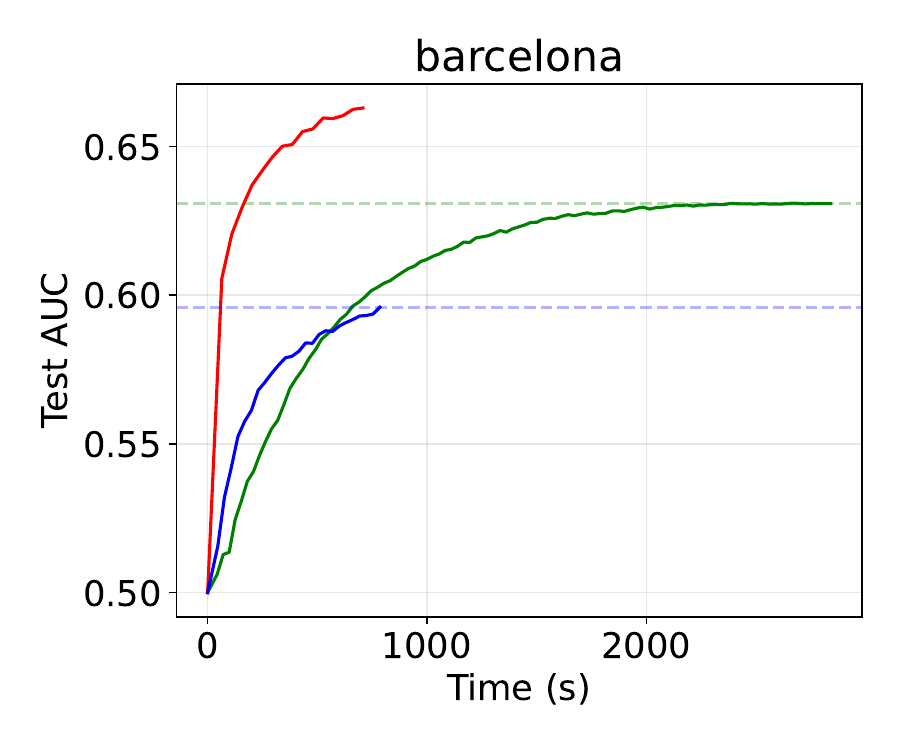}
  \end{subfigure}
  \begin{subfigure}[b]{0.32\textwidth}
    \centering
    \includegraphics[width=\textwidth]{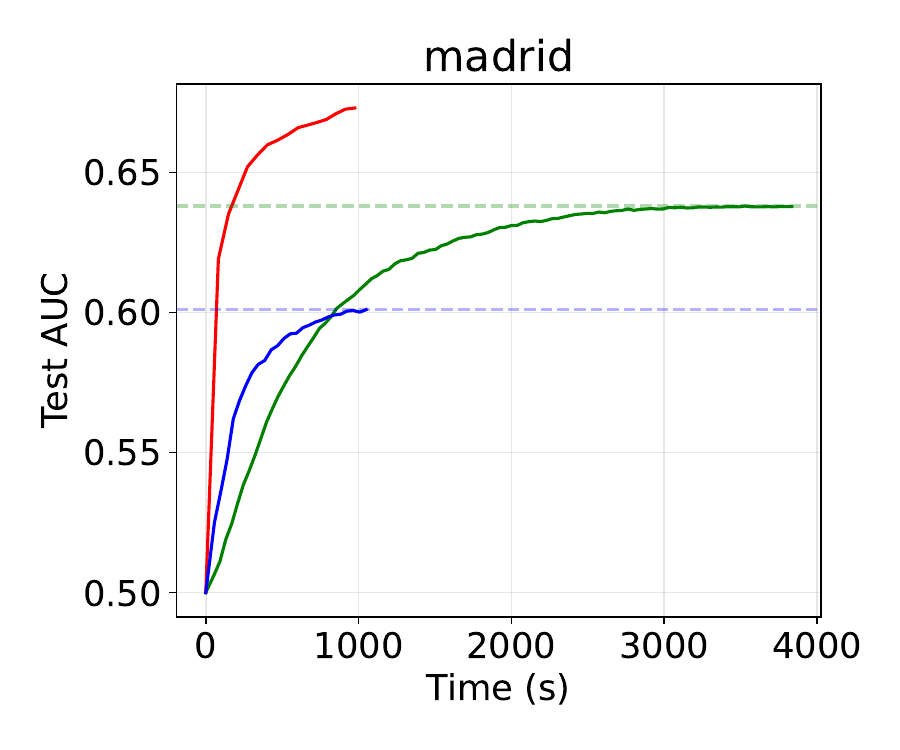}
  \end{subfigure}
    
  \begin{subfigure}[b]{0.32\textwidth}
    \centering
    \includegraphics[width=\textwidth]{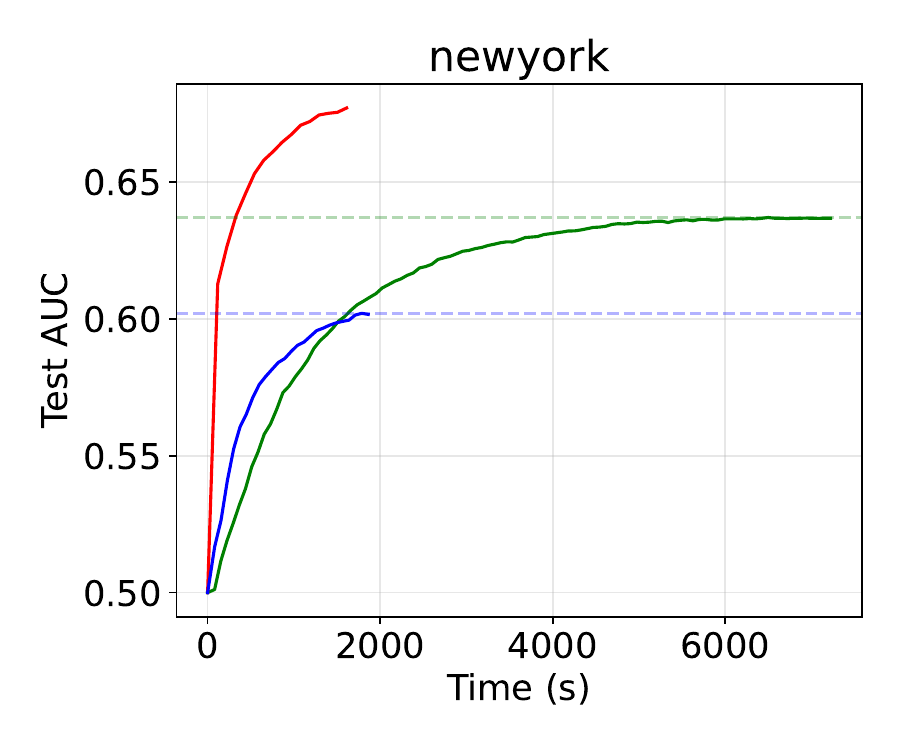}
  \end{subfigure}
  \begin{subfigure}[b]{0.32\textwidth}
    \centering
    \includegraphics[width=\textwidth]{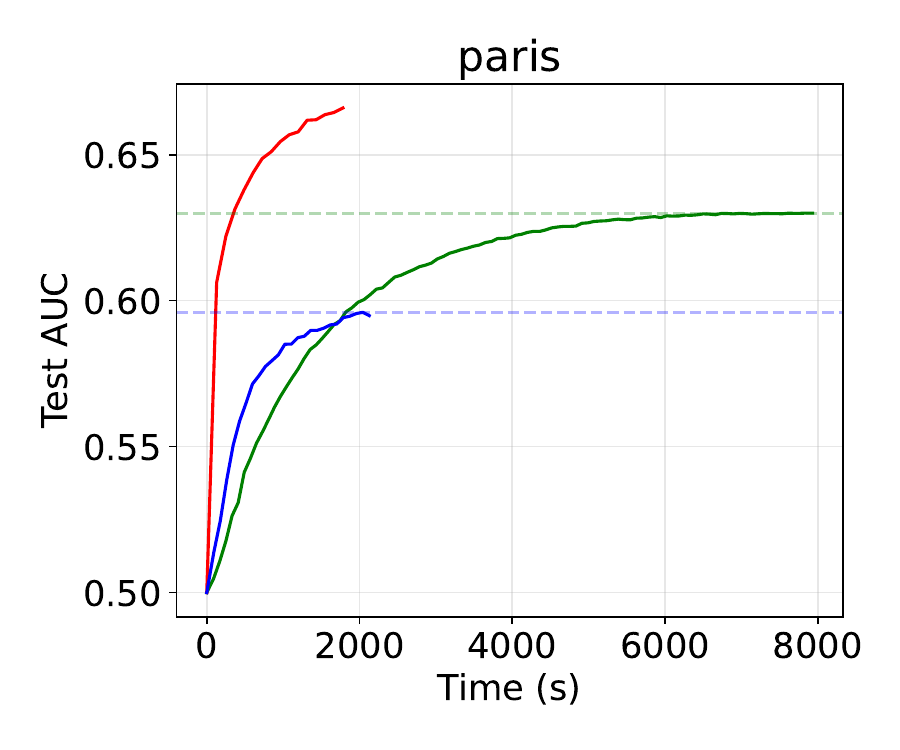}
  \end{subfigure}
  \begin{subfigure}[b]{0.32\textwidth}
    \centering
    \includegraphics[width=\textwidth]{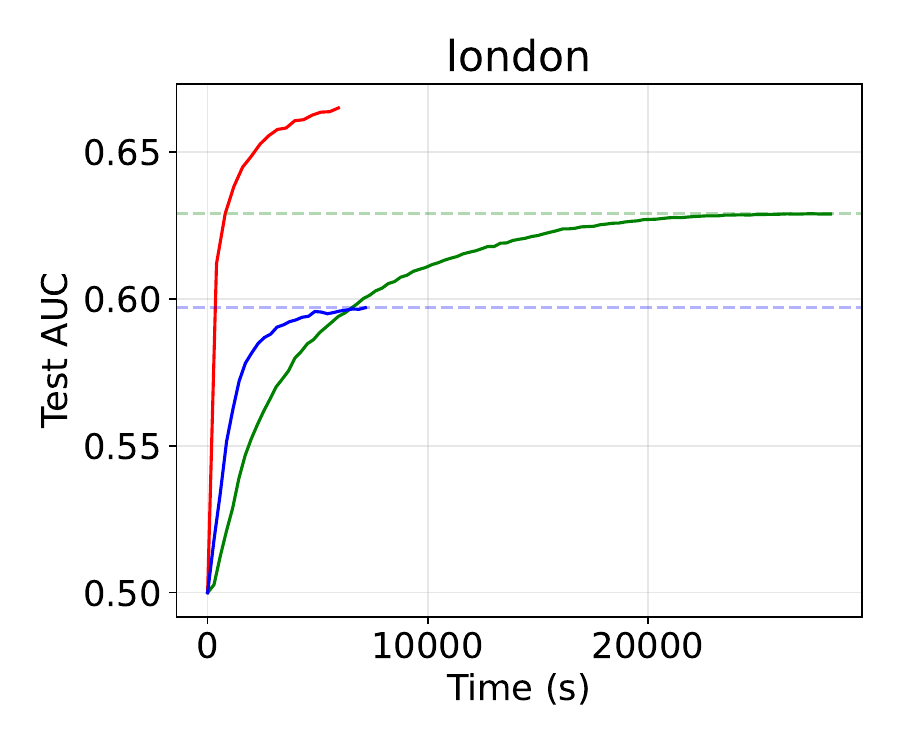}
  \end{subfigure}

    \begin{center}
        \textbf{CO${_2}$ Emissions vs. Test AUC}
    \end{center}
    
  \begin{subfigure}[b]{0.32\textwidth}
    \centering
    \includegraphics[width=\textwidth]{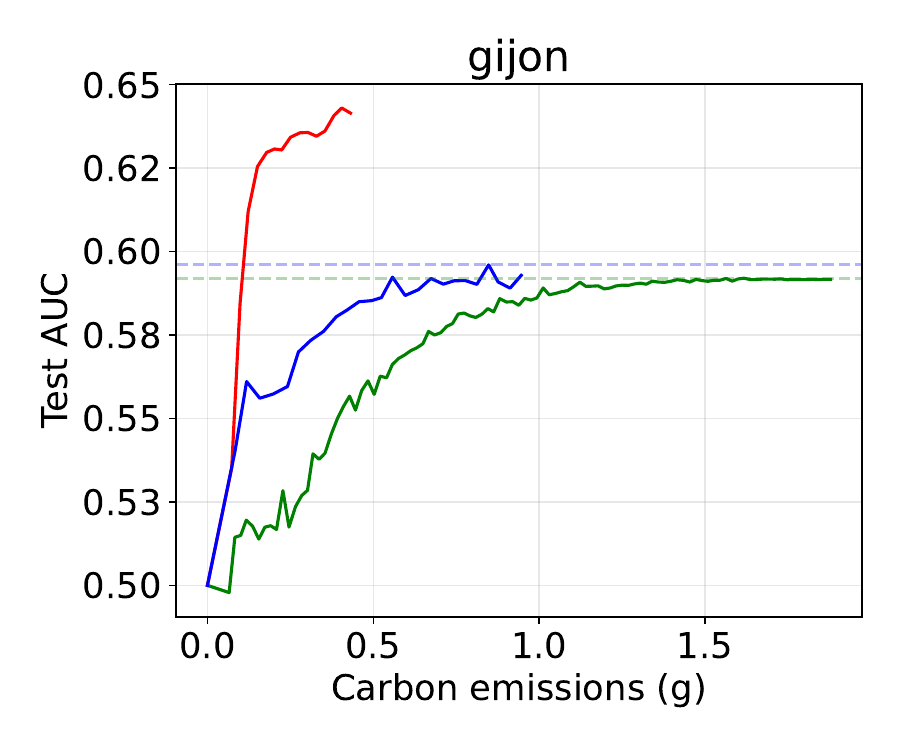}
  \end{subfigure}
  \begin{subfigure}[b]{0.32\textwidth}
    \centering
    \includegraphics[width=\textwidth]{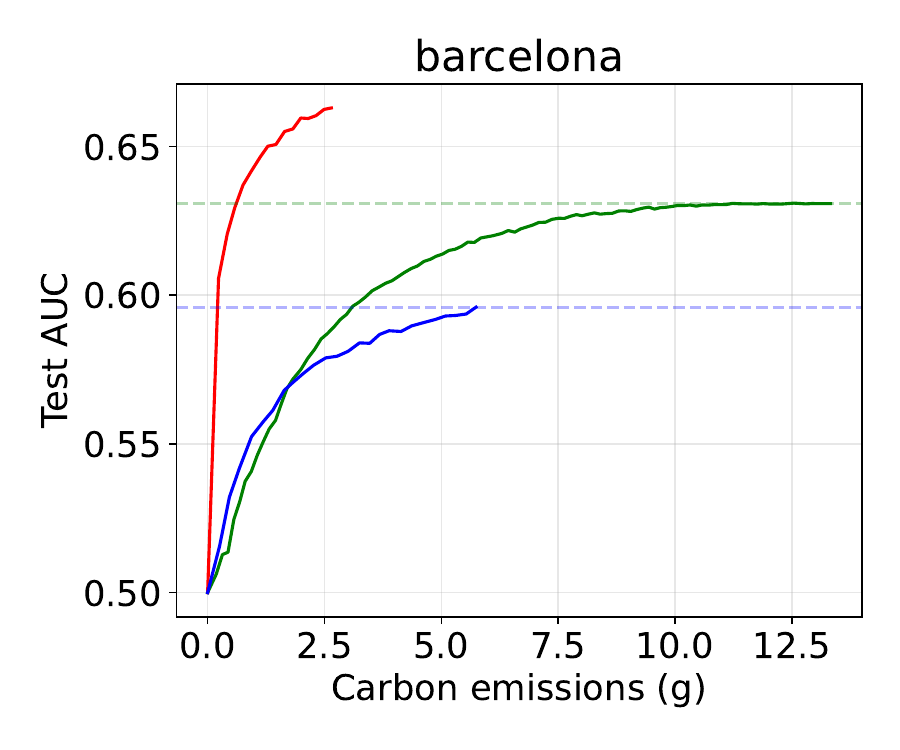}
  \end{subfigure}
  \begin{subfigure}[b]{0.32\textwidth}
    \centering
    \includegraphics[width=\textwidth]{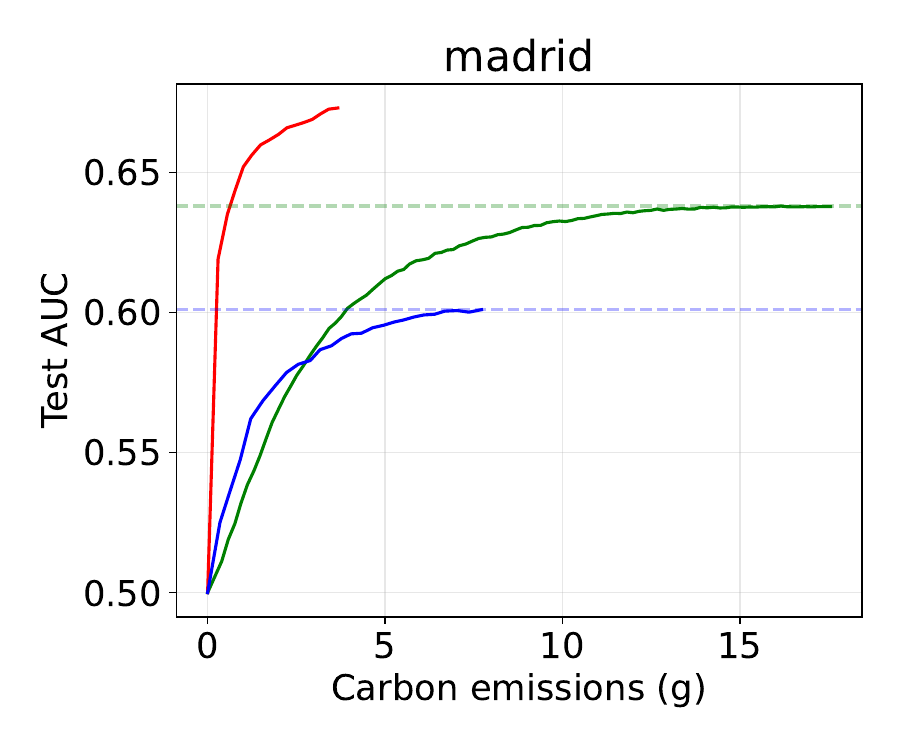}
  \end{subfigure}

  \begin{subfigure}[b]{0.32\textwidth}
    \centering
    \includegraphics[width=\textwidth]{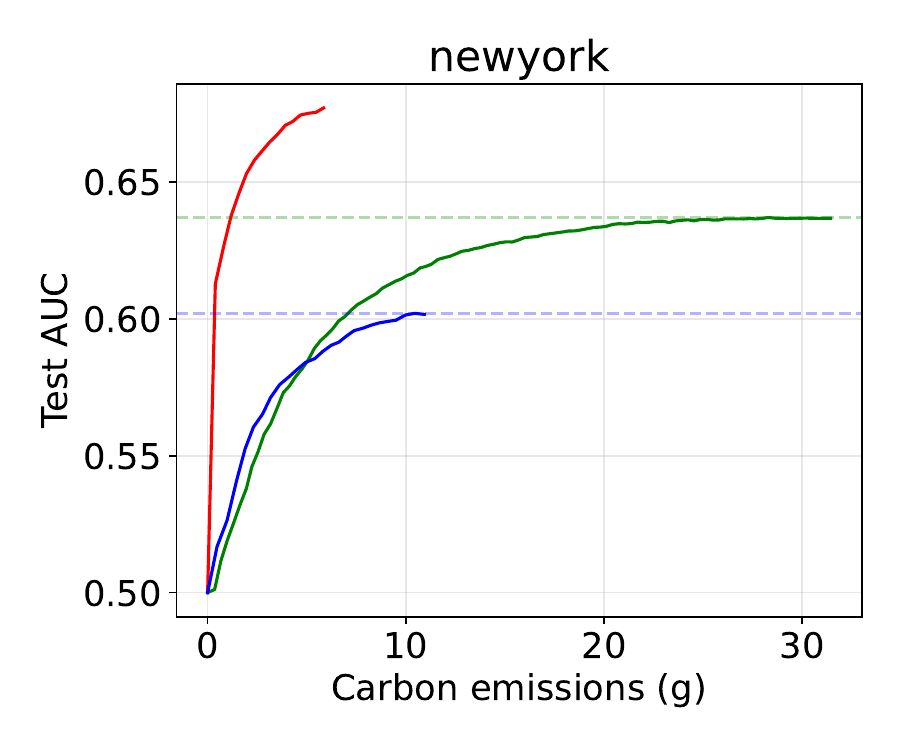}
  \end{subfigure}
  \begin{subfigure}[b]{0.32\textwidth}
    \centering
    \includegraphics[width=\textwidth]{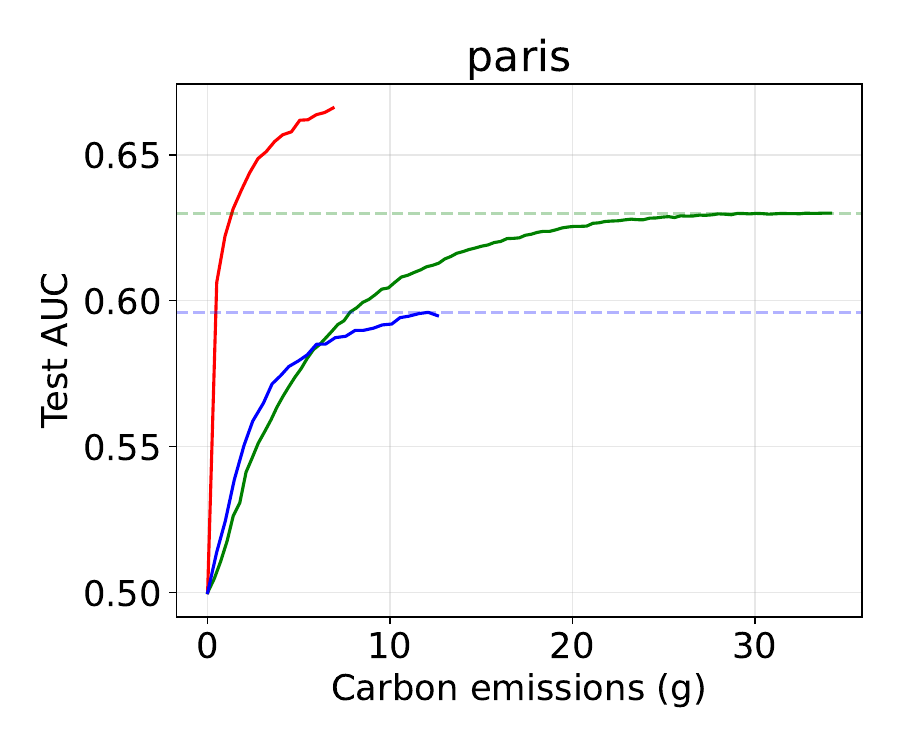}
  \end{subfigure}
  \begin{subfigure}[b]{0.32\textwidth}
    \centering
    \includegraphics[width=\textwidth]{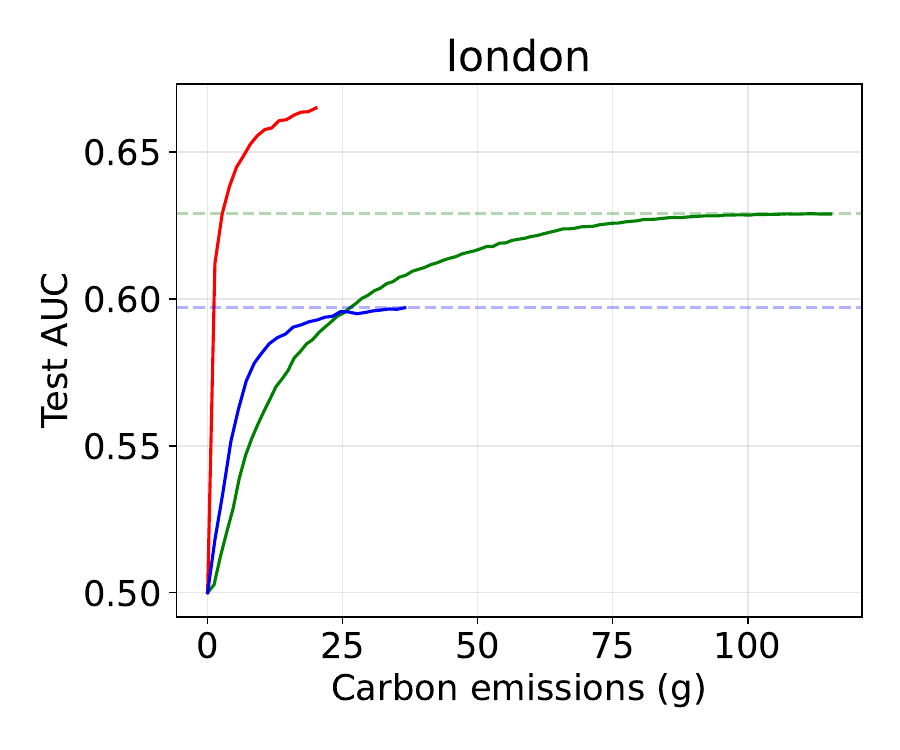}
  \end{subfigure}

    \caption{Comparison of Time vs. AUC (top half) and Carbon Emissions vs. AUC (bottom half) during training for ELVis, MF-ELVis, and BRIE across each city.}

\label{fig:auc_time_carbon}
\end{figure}

Figure \ref{fig:auc_time_carbon} compares AUC evolution versus training time and CO$_2$ emissions for the different models. At the end of the training process, BRIE emits approximately 75\% less CO$_2$ than ELVis and 50\% less than MF-ELVis, the current most sustainable model. Moreover, BRIE requires around three times less training time than ELVis in all cities, tying MF-ELVis as the model with the fastest training. The figure also reveals an interesting insight: BRIE consistently outperforms ELVis and MF-ELVis no matter the training time and emitted CO$_2$, i.e. for any given training time or CO$_2$ emissions goal, BRIE can achieve higher performance compared to the other models. Alternatively, if a specific performance goal is set, BRIE can attain it faster and with lower CO$_2$ emissions.

\begin{figure}[h]
\centering
\includegraphics[width=0.8\textwidth]{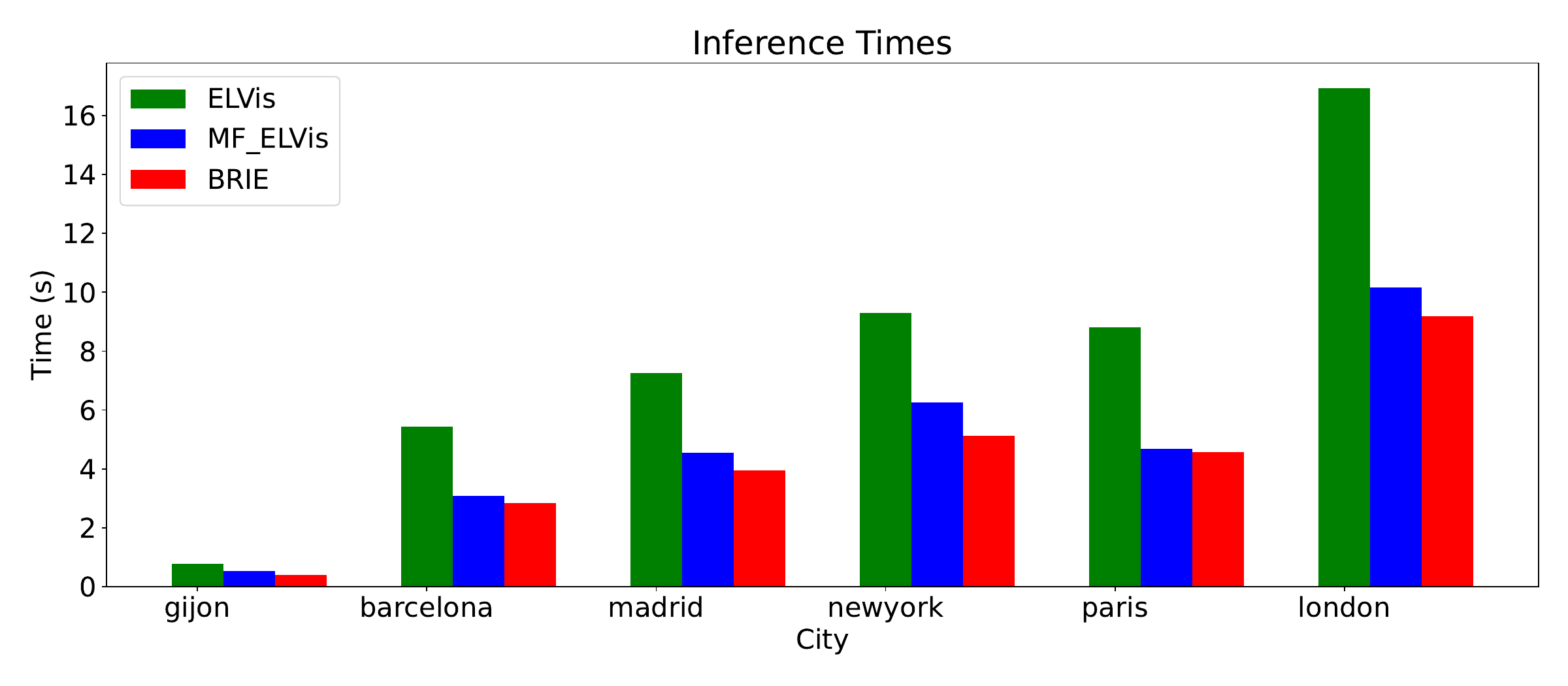}
\vspace{0.5cm}
\includegraphics[width=0.8\textwidth]{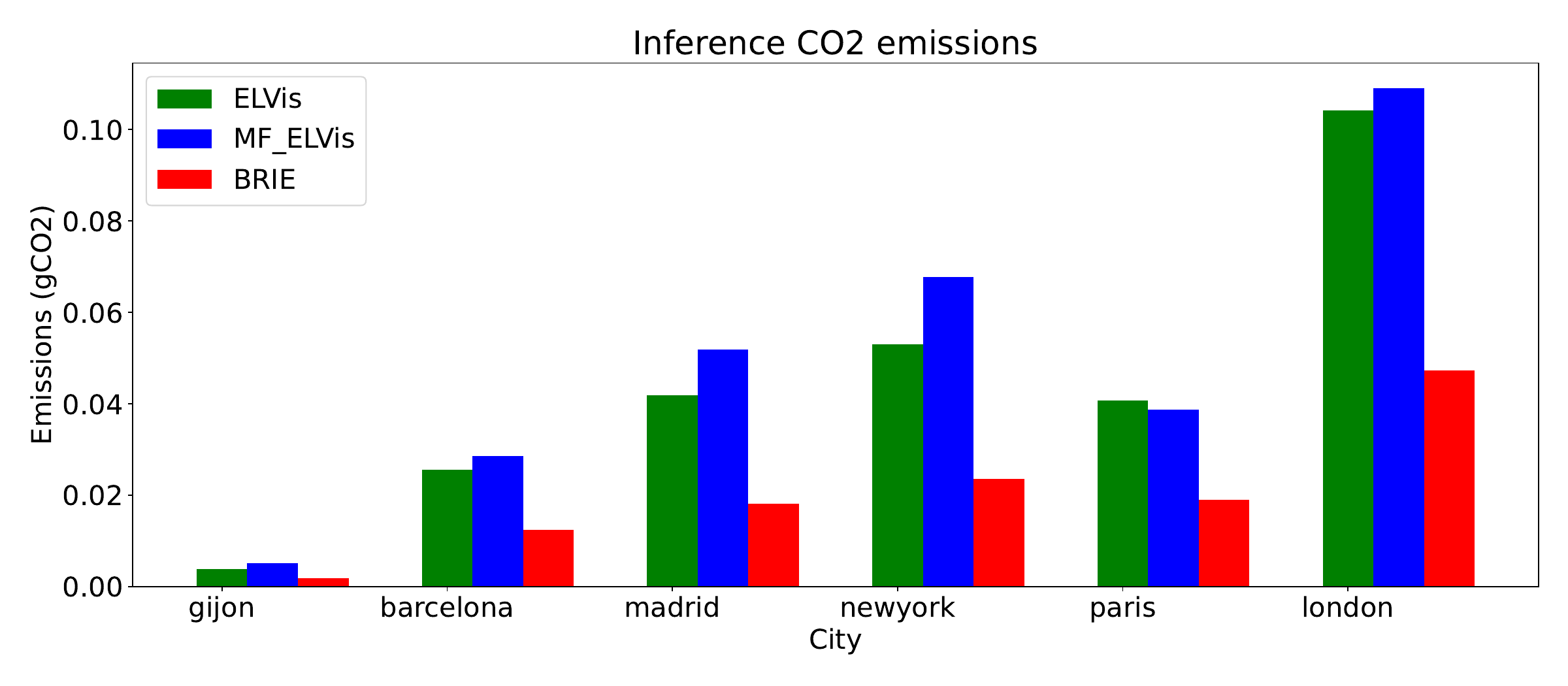}
\caption{Comparison of inference times (top) and inference emissions (bottom) for different models across the six datasets. The results represent the average inference costs of predicting the test set in each dataset, averaged over 50 runs. GPU acceleration using CUDA was employed for both training and inference.}
\label{inference_time_carbon}
\end{figure}

With respect to inference, Figure \ref{inference_time_carbon} shows BRIE and MF-ELVis present no significant variation and are consistently around two times faster than ELVis. Furthermore, BRIE emits 50\% less CO$_2$ during inference compared to both MF-ELVis and ELVis. 

Overall, these findings can be attributed to several factors. Firstly, BRIE's more adequate learning strategy based on Bayesian Pairwise Ranking (BPR) and handling of training data enables it to achieve higher performance early in the training process. Secondly, the complexity and resource requirements of the use of an MLP as a learned similarity function in ELVis incur penalties in terms of training and inference CO$_2$ emissions and time compared to BRIE's fixed similarity function based on dot product.

Moreover, the high embedding dimension (number of factors $d$) employed in MF-ELVis ($d=1024$) and ELVis ($d=256$) leads to a substantial penalty in terms of emissions. Notably, although BRIE requires two different images $p$ and $p_{neg}$ in each training sample instead of only $p$, the training times and emissions for BRIE are comparable to MF-ELVis's, and the inference emissions of MF-ELVis are in some cases higher than those of ELVis. It is also worth noting that the use of CUDA in this experiment mitigates larger penalties associated with the high embedding dimensions of ELVis and MF-ELVis, as these would be highly inefficient to handle using CPU only, in contrast with the smaller embeddings managed by BRIE.  

To summarize, BRIE not only achieves higher performances but also provides higher sustainability than existing state-of-the-art models, establishing a lead early in the training process and a high inference efficiency.

\newpage
\subsection{Compactness comparison}

\begin{figure}[h] 
    \begin{center}
        \textbf{Number of latent factors $\mathbf{d}$ vs. Test AUC}
    \end{center}
  \centering
  \begin{subfigure}{0.32\textwidth}
    \includegraphics[width=\linewidth]{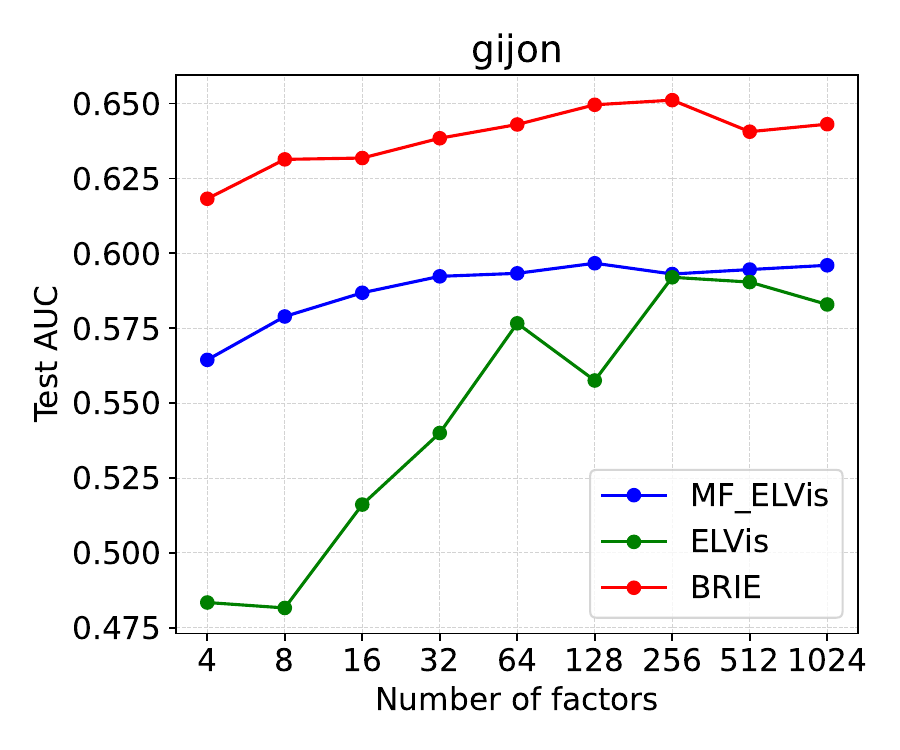}
  \end{subfigure}
  \hfill
  \begin{subfigure}{0.32\textwidth}
    \includegraphics[width=\linewidth]{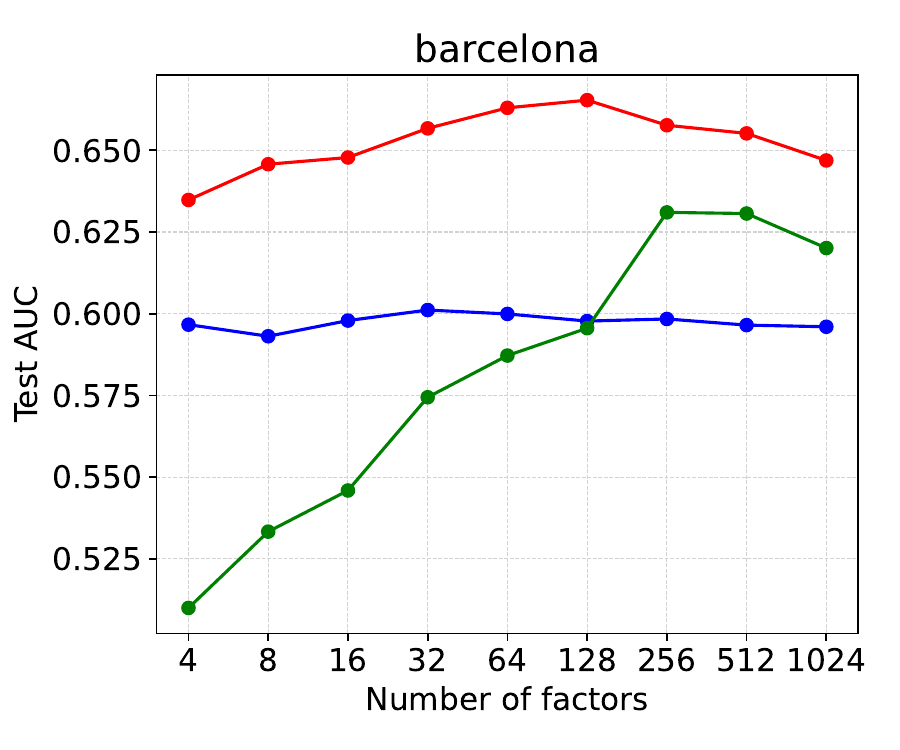}
  \end{subfigure}
  \hfill
  \begin{subfigure}{0.32\textwidth}
    \includegraphics[width=\linewidth]{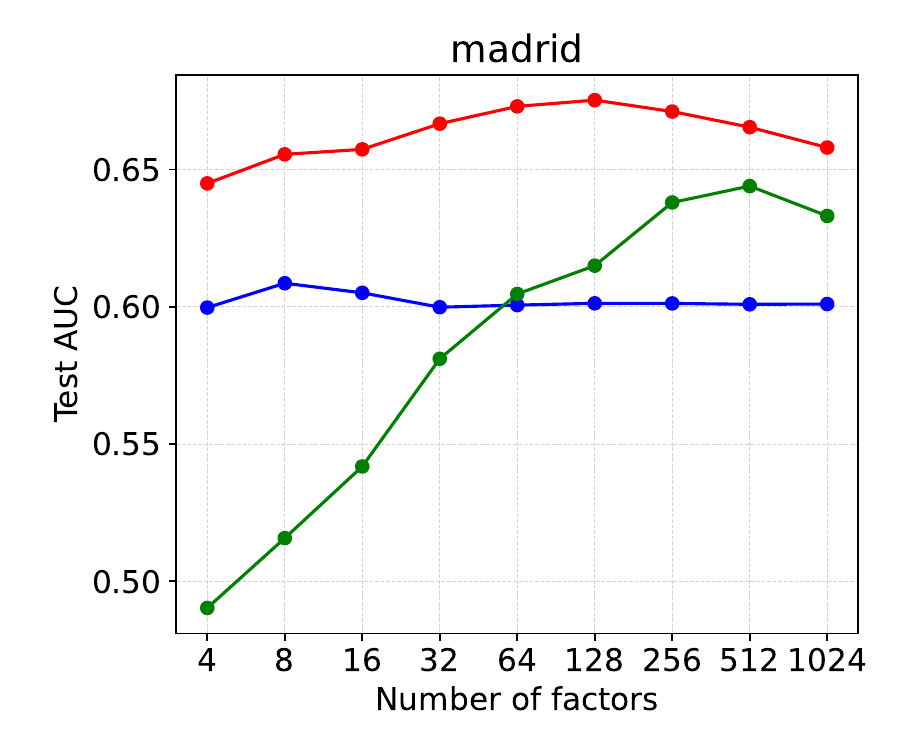}
  \end{subfigure}
  
  \vspace{0.5cm}
  
  \begin{subfigure}{0.32\textwidth}
    \includegraphics[width=\linewidth]{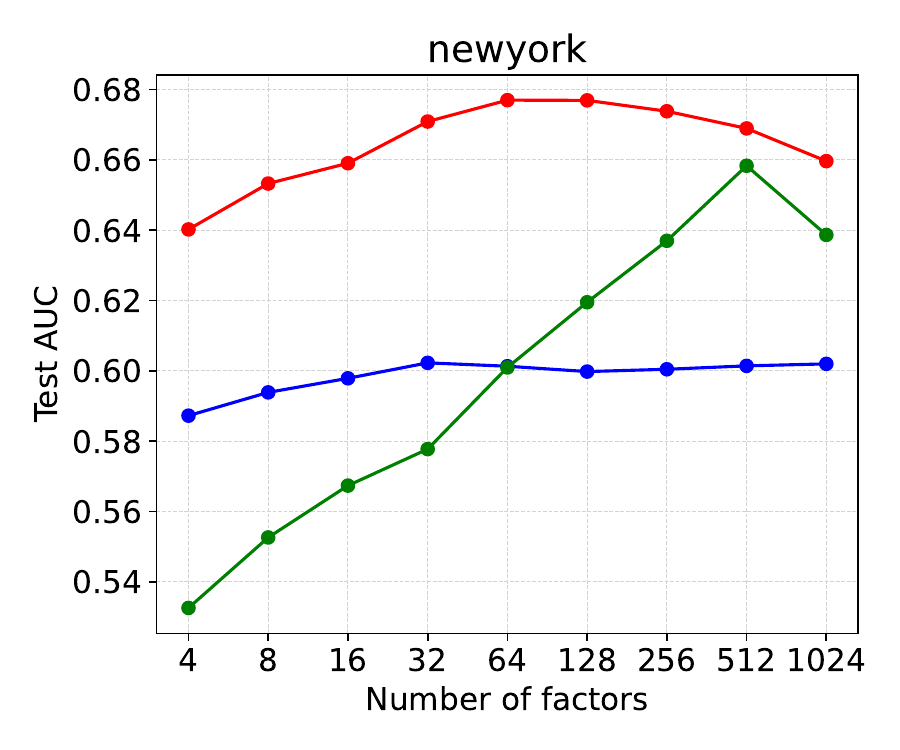}
  \end{subfigure}
  \hfill
  \begin{subfigure}{0.32\textwidth}
    \includegraphics[width=\linewidth]{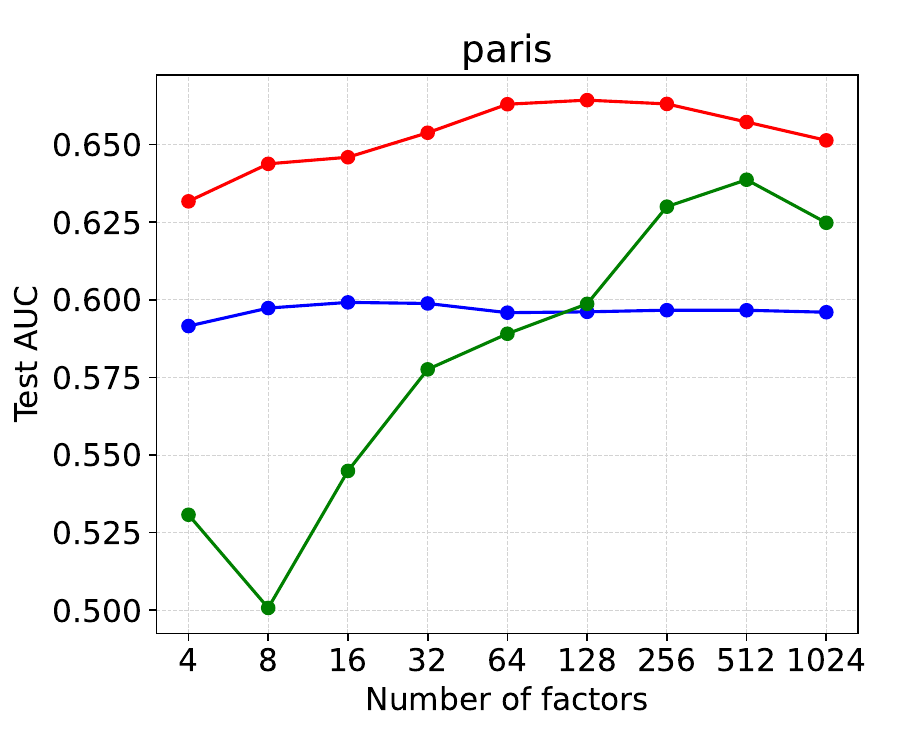}
  \end{subfigure}
  \hfill
  \begin{subfigure}{0.32\textwidth}
    \includegraphics[width=\linewidth]{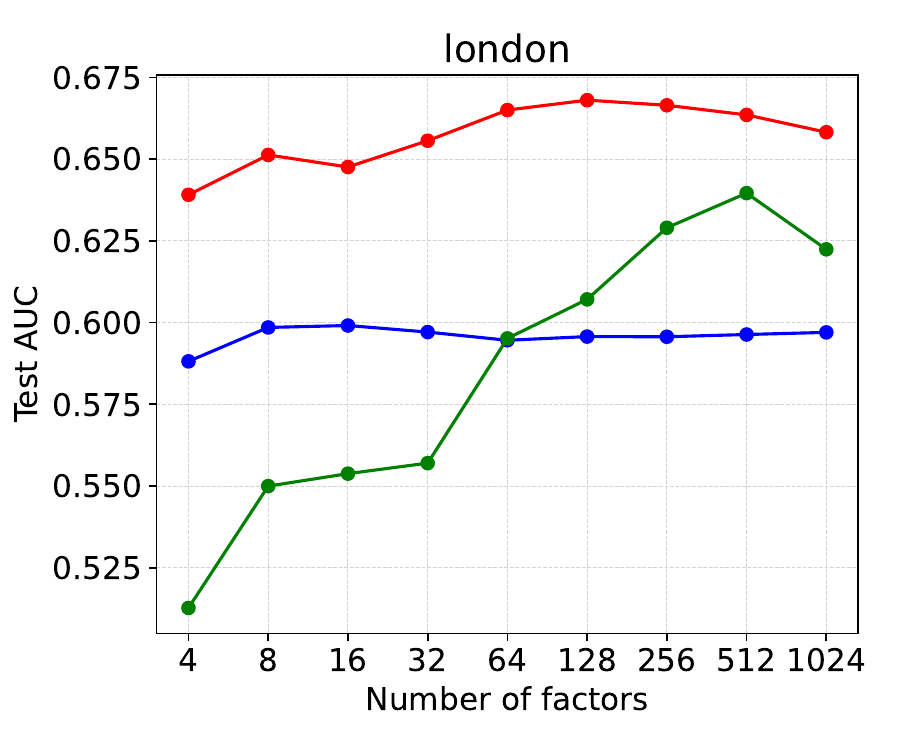}
  \end{subfigure}

  \caption{Impact of the number of factors $d$ (x-axis), the size of the latent embeddings of users and photographs, on the Test MAUC (y-axis) of ELVis, MF-ELVis and BRIE in each dataset.}
  \label{fig:mauc_factors}
\end{figure}

In this section, we examine how the size of the models, particularly the choice of the number of latent factors ($d$), the main contributor to model size, influences their performance. Figure \ref{fig:mauc_factors} presents, for each dataset, the Test AUC values for all three models at varying numbers of latent factors $d$. The results reveal several notable observations across all datasets: 

\begin{itemize}
    \item The performance of MF-ELVis and BRIE remains relatively consistent even with a small number of latent factors.
    \item ELVis experiences a sharp decline in performance as the number of learnt latent factors decreases. In fact, when $d$ is as small as 4, ELVis performs poorly, almost on par with random predictions.
    \item  BRIE consistently outperforms MF-ELVis regardless of the number of latent factors used, and achieves generally better results than ELVis even with a low $d$. In some cases, BRIE with $d=4$ can even outperform ELVis with $d=1024$.
\end{itemize}

The results suggest that MF-ELVis and BRIE demonstrate a higher level of compactness compared to ELVis. The ability of MF-ELVis and BRIE to perform well with a smaller number of latent factors can probably be attributed to the choice of a fixed similarity function in order to compute the suitability of the photograph as an explanation for the user. In contrast, ELVis, which relies on a learned MLP as the similarity function, appears to be more sensitive to the reduction in the number of latent factors, leading to a significant performance drop, likely due to the degradation of information through the series of dropout, activation and dense layers that form the MLP.  

All in all, BRIE offers better compactness and performance robustness in small model sizes when compared to other state-of-the-art methods, making it an attractive option for applications where model size and computational efficiency are of paramount importance.

%% file: 6_Conclusions.tex
\section{Conclusions}
\label{sec:conclusions}

In this work, we have explored the generation of personalized explanations in recommendation systems, specifically focusing on explaining recommendations using user-uploaded photographs. Our approach, BRIE, has shown promising results in addressing the limitations of existing models by leveraging a more adequate learning goal based on Bayesian Pairwise Ranking (BPR) and a more efficient handling of training data. In a state-of-the-art evaluation scheme comprising six real-world datasets of explainability of restaurant recommendations, BRIE achieves performance consistently superior to approaches of the State of The Art and demonstrates remarkable efficiency, emitting significantly fewer carbon emissions during training and inference. Moreover, BRIE maintains its effectiveness even with reduced computational resources and a drastic decrease in the total model size.

We believe that, with the development of a model like BRIE, we provide a double contribution to the broad field of ethical Machine Learning. First, our approach increases the effectiveness of a novel approximation of visual-based explainability, increasing the potential for transparency of systems that deal with user-item interaction data that contain user-images, such as Recommender Systems, e-commerce platforms, or social networks. This effective transparency is an increasingly necessary trait, promoting user trust and fairness and constituting a cornerstone of AI regulations such as the EU's AI Act \cite{hacker2023sustainable, Aiact}. Second, our method exhibits a remarkably strong computational efficiency and remains disentangled from the overarching AI system it explains: not only it doesn't require the generation of images or auxiliary information, as it uses exclusively existing, user-uploaded images to explain outputs, but it also heavily improves the required time, CO$_2$ emissions and model sizes required to achieve this approach when compared to state-of-the-art models. This quest for more sustainable Artificial Intelligence or Green AI is also a crucial aspect of AI-specific regulations like the aforementioned AI Act, which obliges high-impact general purpose models such as Recommender Systems to measure and minimize their environmental impact \cite{hacker2023sustainable, Aiact}. This way, in the context of Recommender Systems, BRIE constitutes a double step forward: it both enhances the transparency of Recommender Systems, and reduces the environmental impact needed to do so. 

Moving forward, there are important avenues for future work. Firstly, we can explore Positive Unlabelled Learning methods for improved negative sample selection during training to enhance the model's ability to discriminate between positive and negative instances; selecting only ``reliable negatives'' for training has the potential to lead to improved performance and more reliable explanations. Secondly, addressing the ``cold start'' problem is crucial, enabling BRIE to provide personalized explanations to new users efficiently. Techniques such as transfer learning can be explored for this purpose, as currently the system needs to be partially re-trained to provide explanations to new users. Lastly, ensuring the scalability of BRIE is important as recommendation systems deal with increasingly larger datasets and user bases, and the model sizes of BRIE and other models scale linearly with the number of unique users. Investigating methods to optimize the model's scalability and resource utilization will enhance its practical applicability.

%% file: 7_Acknowledgements.tex
This research work has been funded by MICIU/AEI/10.13039/501100011033 and ESF+ (grant FPU21/05783), 
\textit{ERDF A way of making Europe} (PID2019-109238GB-C22) and ERDF/EU (PID2021-128045OA-I00), and by the Xunta de Galicia (Grant ED431C 2022/44) with the European Union ERDF funds. CITIC, as Research Center accredited by Galician University System, is funded by ``Consellería de Cultura, Educación e Universidade from Xunta de Galicia'', supported in an 80\% through ERDF Operational Programme Galicia 2014-2020, and the remaining 20\% by ``Secretaría Xeral de Universidades'' (Grant
ED431G 2019/01).

%% file: A_Appendix.tex
\section{On the use of Dropout regularizations in BRIE}

\begin{figure}[h]
    \centering
    \includegraphics[width=.48\textwidth]{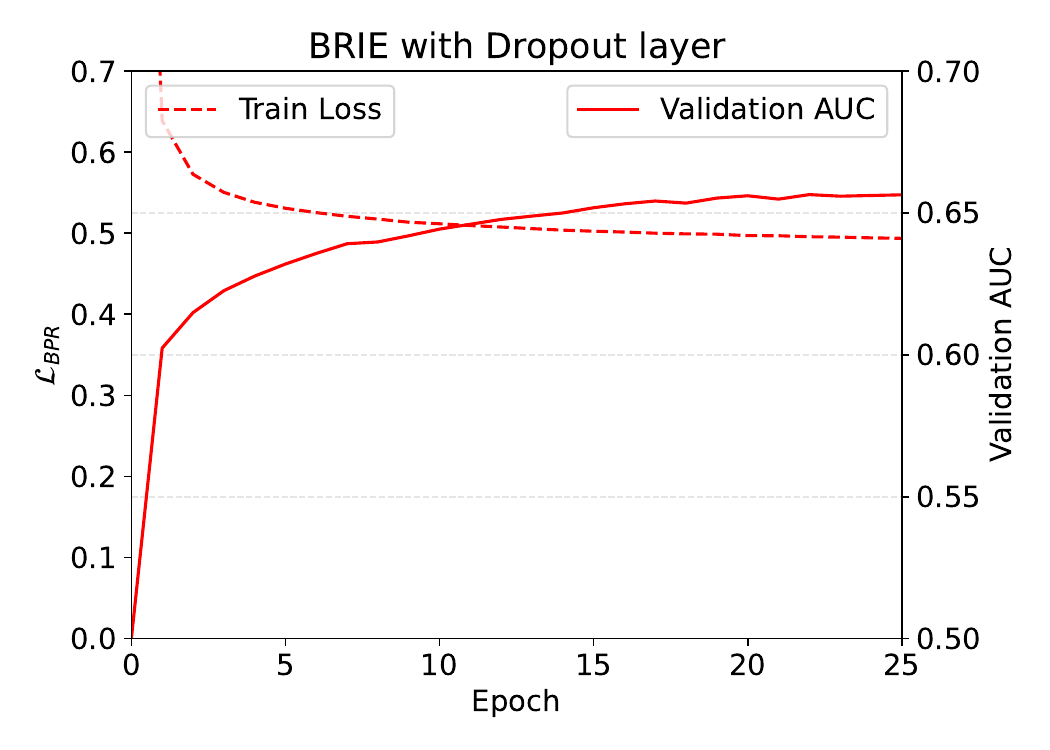}
    \includegraphics[width=.48\textwidth]{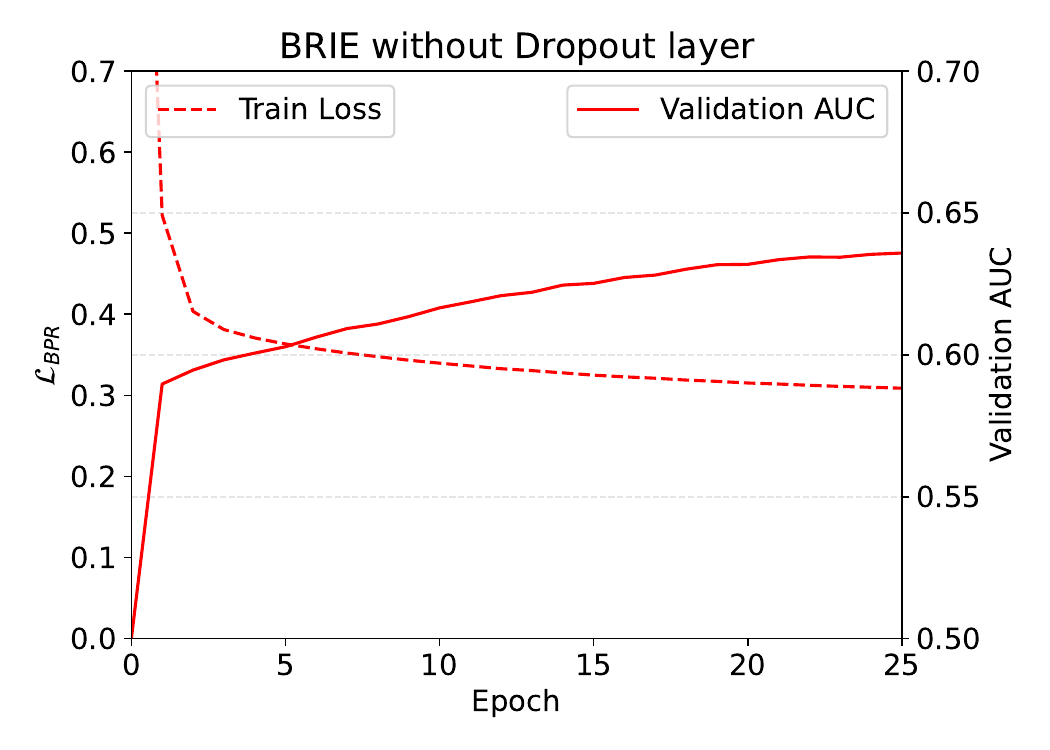}
    
    \caption{Comparison of overfitting in the training process (based on the train loss and validation AUC) of two setups of BRIE. One (left) uses the best found hyperparameter combination described in Section \ref{sec:experimentalsetup}, while the other (right) removes the dropout regularization from that same configuration.}
    \label{fig:dropout}
\end{figure}

In Section \ref{sec:proposal}, we discussed how, to prevent a empirically observed overfitting to training data by BRIE, we decided to apply a dropout regularization to both $U_u$ and $V'_p$, the latent embeddings, to stabilize the learning process and further increase the performance of the model. This overfitting, which appears despite using a low number of latent factors $d$, is most likely caused due to the better suitability of the training policy used by BRIE for the image explanations ranking task. Figure \ref{fig:dropout} shows the evolution of Validation AUC and train loss $\mathcal{L}_{BPR}$ of a fitting process in the dataset used for hyperparameter searching, Barcelona, for two versions of BRIE. The left results correspond to a BRIE that uses the best hyperparameter combinations found in Section \ref{sec:experimentalsetup} ($d$=64, lr=$10^{-3}$, dropout $p=0.75$, 15 epochs), while the right results correspond to a BRIE that removes the dropout layer (i.e. $p=0$), maintaining the rest of the configuration. As it can be observed, while using no dropout layer allows a much tighter fit to the training data. corresponding to a lower train loss, the final performance of BRIE trained with a high dropout is considerably higher, without sacrificing the sustainability of the training process. 

%% file: BRIE.bbl
\begin{thebibliography}{10}
\bibitem{fikret2023transparent}
E.~Fikret, Transparent ml systems for the process industry: How can a recommendation system perceived as transparent be designed for experts in the process industry? (2023).
\end{thebibliography}
